\documentclass{jfm}
\usepackage{graphicx}
\usepackage{epstopdf, epsfig}

\usepackage{amsmath}
\usepackage{amssymb}
\usepackage{xcolor}
\usepackage{mathtools}
\newcommand{\dpr}[2] {\frac{\partial{#1}}{\partial {#2}}}

\usepackage{lineno}

\newcount\ndots
\def\drwln#1#2{\raise 2.5pt\vbox{\hrule width #1pt height #2pt}}
\def\spc#1{\hskip #1pt}
\def\solid{\drwln{24}{1.0}\ }
\def\solidd{\drwln{24}{.1}\ }
\def\chndot{\hbox {\drwln{5}{1.0}\spc{2}\drwln{2}{1.0}
    \spc{2}\drwln{5}{1.0}}\nobreak\ }
\def\dashed{\hbox {\drwln{4}{1.0}\spc{2}
    \drwln{4}{1.0}\spc{2}\drwln{4}{1.0}}\nobreak\ }

\def\dotted{\hbox {\drwln{2}{1.0}\spc{2}
    \drwln{2}{1.0}\spc{2}\drwln{2}{1.0}}\nobreak\ }

\definecolor{ao}{rgb}{0.0, 0.5, 0.0}

\shorttitle{Supersonic flows over rough walls}

\shortauthor{M. Aghaei-Jouybari et al.}

\author{Mostafa Aghaei-Jouybari\aff{1}, Junlin Yuan\aff{2}\corresp{\email{junlin@egr.msu.edu}}, Zhaorui Li\aff{3}, \newline
Giles J. Brereton\aff{2}, \and Farhad A. Jaberi\aff{2}
}
\affiliation{\aff{1}Department of Mechanical Engineering, Johns Hopkins University, Baltimore, \newline MD 21218, USA
\aff{2}Department of Mechanical Engineering, Michigan State University, East Lansing, \newline MI 48824, USA
\aff{3}Department  of  Engineering,  Texas  A\&M  University-Corpus  Christi, Corpus Christi, \newline TX 78412, USA
}

\title{Supersonic turbulent flows over sinusoidal rough walls}

\begin{document}
\maketitle

\begin{abstract}
Direct numerical simulations were performed to characterize fully developed supersonic turbulent channel flows over isothermal rough walls. The effect of roughness was incorporated using a  level-set/volume-of-fluid immersed boundary method. Turbulence statistics of five channel flows are compared, including one reference case with both walls smooth and four cases with smooth top walls and bottom walls with two-dimensional (2D) and   three-dimensional (3D) sinusoidal roughnesses. Results reveal a strong dependence of the turbulence on the roughness topography and the associated shock patterns. Specifically, the 2D geometries generate strong oblique shock waves that propagate across the channel  and are reflected back to the rough-wall side. These strong shocks are absent in the smooth-wall channel and are significantly weaker in cases with 3D roughness geometries, replaced by weak shocklets. At the impingement locations of the shocks on the top wall in the {2D roughness cases}, localized augmentations of turbulence shear production are observed. Such regions of augmented production also exist for the 3D cases, at a much weaker level. The oblique shock waves are thought to be responsible for a more significant  entropy generation for cases with 2D surfaces than those with 3D ones, leading to a higher irreversible heat generation and consequently higher temperature values in 2D roughness cases. In the present supersonic channels, the effects of roughness extend beyond the near-wall layer due to the shocks. {This suggests that outer layer similarity may not fully apply to a rough-wall supersonic turbulent flow.
}

\end{abstract}

\section{Introduction}\label{sec:intro_chp5}
The effects of wall roughness on physics, control, and modeling of compressible flows (subsonic, sonic, super- and hypersonic) are not well understood today.  An understanding of these effects is important for flight control and thermal management of high speed vehicles, especially for reentry applications and reusable launch vehicles. In high speed flow studies, roughness is typically modeled with an $isolated$ (e.g. steps, joints, gaps, etc.), or a $distributed$ (e.g. screw threads, surface finishing, and ablation) organized structure.  \citet{Reda02} and \citet{Schneider08} have reviewed the effects of roughness on boundary layer transition, based on experimental wind-tunnel and in-flight test data of flows in supersonic and hypersonic conditions. \citet{Radeztsky99} analyzed the effects of roughness of a characteristic size of 1-$\mu$m (i.e. a typical surface finish) on transitions in swept-wing flows, and \citet{Latin98} investigated effects of roughness on supersonic boundary layers using rough surfaces with an equivalent sandgrain height  $k_s$ of $O(1\text{mm})$, corresponding to $100<k_s^+<600$ (where  superscript $+$ shows normalization in wall units).
Experimental studies of distributed roughness effects on compressible flows, boundary layer transition, and heat transfer include those of \citet{Braslow58, Reshotko04, Ji06} and \citet{Reda08}.

There are ample studies in the literature focusing on the dynamics, modeling, statistics and structures of flows over rough walls in incompressible regime \citep{Nikuradse33e,RaupachAR91, Jimenez04, Mejia-AlvarezC10, VolinoVSF11, TalapatraK12, Yang15, FlackF18a,ThakkarTBS18a, YuanA18, Ma2021, Aghaei-Jouybari2021}. 
Here, however, we focus  on the compressible (mostly supersonic) flows over rough walls, the understanding of which is much limited. Some of these studies are summarized below.

\citet{Tyson13} analyzed supersonic channel flows over 2D sinusoidal roughness  at Mach number ($M$) of $M=0.3$, $1.5$ and $3$ to understand compressibility effects on mean and turbulence properties across the channel. They used body-fitted grids to perform the simulations and found that the values of velocity deficit decrease with increasing  Mach number. Their results suggest strong alternation of mean and turbulence statistics due to the shock patterns associated with the  roughness.

\citet{Ekoto08} experimentally investigated the effects of square and diamond roughness elements on the supersonic turbulent boundary layers,  to understand how roughness topography alters the local strain-rate distortion, $d_{max}$, which has a direct effect on turbulence production. Their results indicated that the surface with d-type square roughness generate weak bow shocks upstream of the cube elements, causing  small  $d_{max}$ ($\approx-0.01$), and the surface with diamond elements generate strong oblique shocks and expansion waves near the elements, causing a large variation in $d_{max}$ (ranging from $-0.3$ to $0.4$ across the elements). Studies of \citet{Latin98}, \citet{Latin00} and \citet{Latin02} included a comprehensive investigation on supersonic turbulent boundary layers over rough walls. Five rough surfaces (including a 2D bar, a 3D cube, and three different sandgrain roughness) have been analyzed at $M=2.9$. Effects of wall roughness on mean flow, turbulence, energy spectra and flow structures were studied. 
Their results showed strong linear dependence of turbulence statistics on the surface roughness, and also, strong dependencies of turbulent  length scales and inclination angle of coherent motions on the roughness topography.  
\citet{Muppidi12} analyzed the role of ideal distributed roughness on the transition to turbulence in supersonic boundary layers. They  found that counter-rotating vortices, generated by the roughness elements, break the overhead shear layer, leading to an earlier transition to turbulence than on a smooth wall. A similar study was conducted by \citet{Bernardini12}, who investigated the role of isolated cubical roughness on boundary layer transition. Their results suggest that the interaction between hairpin structures, shed by the roughness element, and the shear layer expedites transition to turbulence, regardless of the Mach number.
Recently, \cite{Modesti22} analyzed compressible channel flows over cubical rough wall, both in transitionally and fully rough regimes, and found that the logarithmic velocity deficit (with respect to the baseline smooth wall) depends strongly on the Mach number, and one must employ compressibility transformations to account for the compressibility effects and to establish an appropriate analogy between compressible and incompressible regimes.

Despite   findings in these studies, comprehensive understanding of turbulence statics and dynamics, as well as  their connection to  roughness geometry (especially for distributed roughnesses) and the associated compressibility effects remained unavailable.  
Most numerical studies on this topic have focused on isolated roughness \citet[see e.g.][]{Bernardini12} or ideal distributed roughness such as wavy walls \citet[see e.g.][]{Tyson13}, due to the simplicity in mesh generation and numerical procedures. However, complex distributed roughness is of primary importance and more relevant to flight vehicles, since in high-speed flows ``even the most well-controlled surface will appear rough as the viscous scale becomes sufficiently small" \citep{Marusic10}. Also, according to \citet{Schneider08}, real vehicles may develop surface roughness during the flight which is not present before launch. This flight-induced roughness may be discrete steps and gaps on surfaces from thermal expansion, or distributed roughness induced by ablation or the impact of dust, water, or ice droplets. 

These studies demonstrate the need for better understanding the effects of complex distributed rough surfaces. Numerical studies of flow over complex roughness geometries  benefit from the immersed boundary (IB) method (see a detailed review by \citep{MittalI05}), which has multiple advantages compared to the employment of a body conformal grid, primarily in the ease of mesh generation. A summary of numerical methods based on immersed boundaries in the compressible flow literature is given below.

\citet{Ghias07} used ghost cell method to simulate 2D viscous subsonic compressible flows. They imposed Dirichlet BC for velocity ($u$) and temperature ($T$) on the immersed boundaries. The pressure ($P$) at the boundary was obtained using  the equation of state and the value of density ($\rho$) was obtained through  extrapolation. Their method  was second-order accurate, both locally and globally. \citet{Chaudhuri11} used the ghost cell  method  to simulate 2D inviscid, sub- and supersonic compressible flows. They applied direct forcing for $\rho$, $u$ and total energy ($E$) equations, while $P$ was determined based on the equation of state.
They used a fifth-order-accurate  WENO shock-capturing scheme by using two layers of ghost cells. \citet{YuanZ18} also used ghost cell method to simulate 2D (sub- and supersonic) compressible flows around moving bodies.
\citet{Vitturi07} used a discretized forcing approach for a finite volume solver to simulate  2D/3D viscous subsonic multiphase compressible flows; the forcing term was determined based on an interpolation procedure. They imposed Dirichlet  BC for $u$ and $T$; the equation of state was used for $P$ and flux correction for $\rho$ and $E$.
\citet{Wang17} used continuous forcing (or penalty IB method) to simulate fluid-structure interaction with 2D compressible (sub, super, and hyper sonic) multiphase flows. 

Most of these studies used sharp-interface IB methods, which allows the boundary conditions (BCs) at the interface to be imposed exactly.   However, 3D flows with complex interface geometries (especially with moving interfaces) cause difficulties and require special considerations. Specifically, issues arise when there are multiple image points for a ghost cell, or when there is none. \citet{Luo17} addressed some of these issues in 2D domains. In addition, the interpolation schemes are dependent on the ghost point locations in the solid domain; the situation becomes complicated for 3D domains.
To account for these difficulties in 3D flows with complex interface geometries, the boundary values can be imposed through a prescribed distribution  across the interface, instead of being imposed precisely. Examples include the approaches based on fluid-volume-fraction weighting proposed by \citet{FadlunVOM00} and \citet{Scotti06}, developed for incompressible flows.



In this study we first introduce a compressible-flow IB method that is { a combination of both volume-of-fluid (VOF) and level-set methods---a modified version of the level-set methods used in  multi-phase flow simulations in the incompressible regimes \citep{Sussman94, Sussman99} and compressible regimes \citep{Li08}, as well as the VOF method of  \citet{Scotti06} for incompressible rough wall flows}---and validate it by comparing  mean and turbulence statistics with a baseline simulation using a body-fitted mesh. Then we analyze the simulation results in supersonic channel flows at $M=1.5$ and a bulk Reynolds number  of 3000 (based on the channel half height) over two 2D and two 3D sinusoidal surfaces. 
{ Analyses are first carried out  with respect to mean quantities and turbulent statistics for an overall comparison of the flow fields, then with the transport equations of Reynolds stresses to compare turbulence production and transports, and finally we conduct conditional analyses of energy budget attributed to solenoidal,  compression and expansion regions of the flow. }
 Section \ref{sec5:method} describes the governing equations, numerical setup and the IBM formulation. Results of the mean flow and turbulence statistics, Reynolds stress budgets and conditional analysis are explained in section \ref{sec5:results}, and the manuscript is concluded in section \ref{sec5:conclusion}.

\section{Numerical setup and formulation}\label{sec5:method}  
\subsection{Governing equations}\label{subsec:gov}
The non-dimensional forms of compressible Navier-Stokes equations are 

\begin{subequations}\label{eq:NS_Compr}
\begin{align}
    &\dpr{\rho}{t}+\dpr{}{x_i}(\rho u_i)=0,\\
    &\dpr{\rho u_i}{t}+\dpr{}{x_j}\big(\rho u_i u_j+p\delta_{ij}-\frac{1}{\textup{Re}}\tau_{ij}\big)=f_1\delta_{i1},\\
    &\dpr{E}{t}+\dpr{}{x_i}\bigg[u_i(E+p) -\frac{1}{\textup{Re}}u_j\tau_{ij}+\frac{1}{(\gamma-1)\textup{Pr}  \textup{Re}M^2}q_i\bigg]=f_1u_1,
\end{align}
\end{subequations}
where $x_1$, $x_2$, $x_3$ (or $x$, $y$, $z$) are coordinates in the streamwise, wall-normal and spanwise directions, with corresponding velocities of $u_1$, $u_2$ and $u_3$ (or $u$, $v$ and $w$). Density, pressure, temperature and dynamic viscosity are denoted by $\rho$, $p$, $T$ and $\mu$, respectively. $E=p/(\gamma-1)+\rho u_iu_i/2$ is the total energy, $\gamma \equiv C_p/C_v$ is the ratio of specific heats (assumed to be 1.4), $\tau_{ij}=\mu\big(\dpr{u_i}{x_j}+\dpr{u_j}{x_i}-\frac{2}{3}\dpr{u_k}{x_k}\delta_{ij}\big)$ is the viscous stress tensor, and $q_i=-\mu \dpr{T}{x_i}$ is the thermal heat flux. $f_1$ is a body force that drives the flow in the streamwise direction, analogous to the pressure gradient.  
The reference Reynolds, Mach and Prandtl numbers are, respectively, $\textup{Re} \equiv \rho_r U_r L_r/\mu_r$, $M\equiv U_r/\sqrt{\gamma R T_r}$, and $\textup{Pr} \equiv \mu_r C_p/\kappa_c$, where subscript $r$ stands for reference values (to be defined in section \ref{sec:rough_para}). The gas constant $R$ and the specific heats $C_p$ and $C_v$ are assumed to be constant throughout the domain (calorically perfect gas). They are related by $R=C_p-C_v$. The heat conductivity coefficient is denoted by $\kappa_c$.

The set of equations in (\ref{eq:NS_Compr}) is closed through the equation of state, which for a perfect gas is 
\begin{equation}\label{eq:state}
p=\frac{\rho T}{\gamma M^2}.    
\end{equation}

Equations (\ref{eq:NS_Compr}) and (\ref{eq:state}) are solved using a finite-difference method in a conservative format and a generalized coordinate system. A fifth-order monotonicity-preserving (MP) shock-capturing scheme and a sixth order compact scheme are utilized for calculating the inviscid and viscous fluxes respectively. The solver uses local Lax-Friedriches (LLF) flux-splitting method and employs an explicit third-order Runge-Kutta scheme for time advancement. The above computational solver or a version of it has been used for direct numerical simulation (DNS) and large-eddy simulations (LES) of a range of compressible turbulent flows including those involving smooth surfaces \citep{Tian17,Tian19,Jammalamadaka13,Jammalamadaka14,Jammalamadaka15}. Readers are referred to \citet{LiJ12} for extensive details of the compressible solver.

\subsection{Details of the present IB method}
The present IB method is a combination of level-set  \citep{Sussman94,Gibou18,Li08} and volume-of-fluid     \citep[VOF,][]{Scotti06} methods. It is designed for  stationary interfaces only. The level-set field $\psi(x,y,z)$  is defined as the signed distance from  the fluid-solid interface. Based on the prescribed roughness geometry, the $\psi$ field is obtained by diffusing an initial discontinuous marker function,
\begin{equation}\label{eq:psi_init}
    \psi_0(x,y,z)=\begin{cases}
      ~~1 & \text{~~~~in fluid cells,}\\
      ~~0 & \text{~~~~in interface cells,}\\
      -1  & \text{~~~~in solid cells,}
    \end{cases}    
\end{equation}
 in the interface-normal direction until a narrow band along the interface, within which $\psi$ is sign-distanced, is generated; this is  similar to the reinitialization process conducted by  \citet{Sussman94}, and done by solving 
\begin{equation}\label{eq:psi}
    \frac{\partial \psi}{\partial \tau}=\text{sign}(\psi)(1-\vert \boldsymbol{\nabla} \psi \vert),
\end{equation}
where $\tau$ is a fictitious time controlling the width of the interface band. 
It is sufficient to march in (fictitious) time until a band width  of up to 2-3 grid size is obtained. 
 
Based on the level-set field,  the VOF field, $\phi(x,y,z)$, is constructed  as 
 \begin{equation}\label{eq:phi}
     \phi\equiv(1+\psi)/2,
 \end{equation}
 such that  $\phi=0$, $0<\phi<1$, and $\phi=1$ correspond to the  solid, interface and fluid cells, respectively. 
 
 To impose the desired boundary condition for a test variable $\theta(x,y,z,t)$, we   correct the values of the variable at the beginning of each Runge-Kutta  substep. The correction is similar to the approach used by ~\citet{Scotti06} and that of Yuan and co-workers~\citep{YuanP14b,YuanYP15,YuanYMBIV19a,Shen20,Mangavelli21}, i.e.,
 \begin{equation}\label{eq:DritBC}
     \theta\rightarrow\phi \theta + (1-\phi)\theta_b,
 \end{equation}
 for Dirichlet BC  and 
 \begin{equation}\label{eq:NeuBC}
     \dpr{\theta}{n}=\boldsymbol{\nabla}\theta\cdot \boldsymbol{\widehat{n}}=\dpr{\theta}{n}\bigg\vert_b
 \end{equation}
 for Neumann BC, where the subscript $b$  denotes   boundary values and $ \boldsymbol{\widehat{n}}$ is the unit normal vector pointing into the fluid  region at the interface.  $ \boldsymbol{\widehat{n}}$ is obtained as 
\begin{equation}\label{eq:ndir}
    \boldsymbol{\widehat{n}}=\boldsymbol{\nabla} \psi =\boldsymbol{\nabla} \phi/\vert \boldsymbol{\nabla} \phi \vert.
\end{equation}
Note that $\phi(x,y,z)$ does not  represent exactly the fluid volume fraction in each grid cell. Instead, $\phi$ is termed the VOF field  because of the analogy between the BC imposition in equations  (\ref{eq:DritBC}) and (\ref{eq:NeuBC}) and the approach  of \citet{Scotti06} using the exact volume-of-fluid.  
As will be shown in section~\ref{sec:valid}, the accuracy of the IB method herein is sufficient to produce matching single-point statistics compared to a simulation using body-fitted grid. 

\subsection{Surface roughnesses and simulation parameters}\label{sec:rough_para}

\begin{figure}
   \centerline{\includegraphics[width=.85\textwidth,trim={0 0cm 0 0cm},clip]{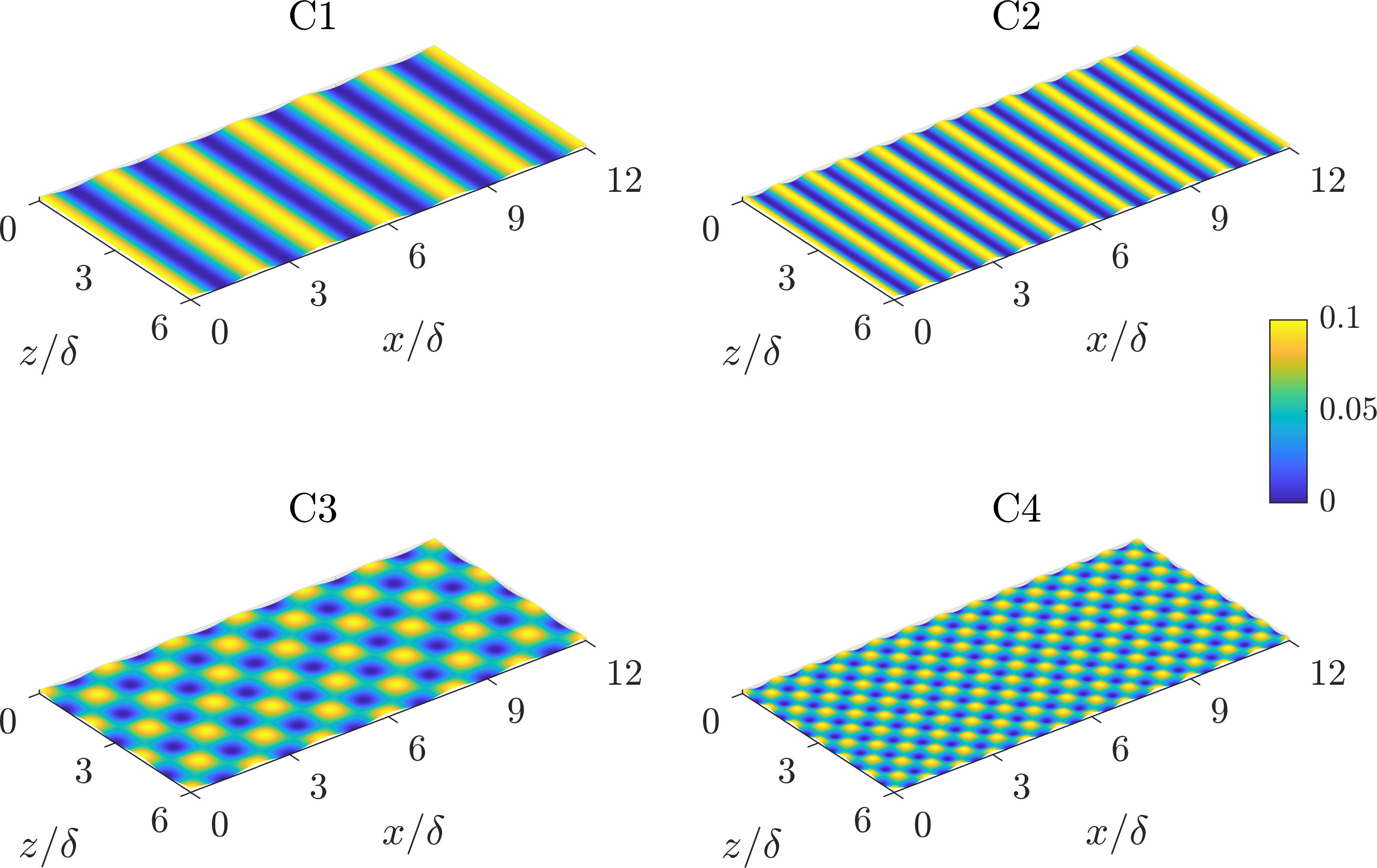}}   
  \caption{Surface roughnesses.}
\label{fig5:roughness}
\end{figure}

Fully developed, periodic compressible channel flows are simulated using four roughness topographies. The channels are roughened only at one surface (bottom wall) and the other surface is smooth. A reference smooth-wall channel is also simulated for validation and comparison purposes. The channel dimensions in streamwise, wall-normal and spanwise directions are, respectively, $L_x=12\delta$, $L_y=2\delta$ and $L_z=6\delta$, where $\delta$ is the channel half-height. 

Figure \ref{fig5:roughness} shows four roughness topographies used for the present simulations. These rough-wall cases are C1-C4, and the smooth-wall baseline case is denoted as SM. All rough cases share the same crest height, $k_c=0.1\delta$. The trough location is set at $y=0$. Cases C1 and C2 are 2D sinusoidal surfaces with streamwise wave-lengths of $\lambda_x=2\delta$ and  $\lambda_x=\delta$, respectively. The roughness heights, $k(x,z)$, for these surfaces are prescribed as
\begin{equation}\label{eq:hei2d}
    k(x,z)/\delta=0.05\big [1+\cos(2\pi x/\lambda_x)\big].
\end{equation}
Cases C3 and C4 are 3D sinusoidal surfaces with equal streamwise and spanwise wave-lengths of $(\lambda_x, \lambda_z)=(2\delta,2\delta)$ for C3, and  $(\lambda_x, \lambda_z)=(\delta,\delta)$ for C4. The roughness heights for them are prescribed as 
\begin{equation}\label{eq:hei3d}
    k(x,z)/\delta=0.05\big [1+\cos(2\pi x/\lambda_x)\cos(2\pi z/\lambda_z)\big].
\end{equation}
Table \ref{tab5:roughness} summarizes some statistical properties of the surface geometries. These statistics are various moments of surface height, surface effective slopes and porosity. 

 \begin{table}
   \begin{center}{\footnotesize 
     \def~{\hphantom{10}} 
     \begin{tabular}{l c c c c c c c c c c}
  Case	  &   $k_c$	 & 	$k_{avg}$	 &  	$k_{rms}$	 & 	$R_a$	 	 & 	$E_x$	 & 	 $E_z$	 & 	 $S_k$ & 	 $K_u$  \\
C1	 & 0.1	 & 	0.05	 & 	0.035 & 0.032		 & 	 	0.100	 & 	0.000	 & 		0.0 & 1.50		 	\\
C2	 & 0.1	 & 	0.05	 & 	0.035 & 0.032		 & 	 	0.200	 & 	0.000	 & 		0.0 & 1.50		 	\\
C3	 & 0.1	 & 	0.05	 & 	0.025 & 0.020		 & 		0.064	 & 	0.064	 & 		0.0 & 2.25		 	\\
C4	 & 0.1	 & 	0.05	 & 	0.025 & 0.020		 & 	 	0.127	 & 	0.127	 & 		0.0 & 2.25		 	\\
     \end{tabular}
     




     \caption{Statistical parameters of roughness topography. $k_c$ is the peak-to-trough height, $k_{avg} = \frac{1}{A_t}\int_{x,z}k(x,z)dA$ is the average height,
     $k_{rms} = \sqrt{\frac{1}{A_t}\int_{x,z} (k-k_{avg})^2 dA}$ is the root-mean-square (r.m.s.) of roughness height fluctuation,
     $R_a = \frac{1}{A_t}\int_{x,z}\vert k - k_{avg}\vert dA$ is the first-order moment of height fluctuations, 
     $E_{x_i} = \frac{1}{A_t}\int_{x,z}\Big|\frac{\partial k}{\partial x_i} \Big|dA$ is the effective slope in the $x_i$ direction,
     $ S_k = \left.\frac{1}{A_t}\int_{x,z}(k - k_{avg})^3dA \right/ k_{rms}^3$ is the height skewness, and
     $ K_u = \left.\frac{1}{A_t}\int_{x,z}(k - k_{avg})^4dA\right/ k_{rms}^4$ is the height kurtosis. Here, $k(x,z)$ is the roughness height distribution; $A_t$ the total planar areas of channel wall. Values of $k_c$, $k_{avg}$, $k_{rms}$ and $R_a$ are normalized by $\delta$.}
   \label{tab5:roughness}}
   \end{center}
 \end{table}

For a test variable $\theta$, the time, Favre and spatial averaging operators are denoted respectively by $\overline{\theta}$, $\widetilde{\theta}=\overline{\rho\theta}/\overline{\rho}$ and $\big<\theta\big>$.
{ Intrinsic planar averaging is used for the spatial averaging, where a $y$-dependent fluid variable is averaged per unit fluid
planar area, $\big<\theta\big>=1/A_f\int_{A_f} \theta \hbox{d}A$; $A_f(y)$ is the area of fluid at a given $y$.} { At $y=0$, $A_f=0$ since all area is inside solid. As a result, data at $y=0$ are not included in intrinsically averaged wall-normal profiles. } Fluctuation components $\theta'$, $\theta''$ and $\theta'''$ are defined following the triple decomposition of \citet{RaupachS82}, such that 
\begin{equation}\label{eq:avgOper}
\begin{aligned}
    \theta &= \overline{\theta}+\theta'\\
           &= \widetilde{\theta}+\theta''\\
           &= \big<\overline{\theta}\big>+\theta'''.
\end{aligned}
\end{equation}

\begin{figure}
   \centerline{\includegraphics[width=1\textwidth,trim={0 0cm 0 0cm},clip]{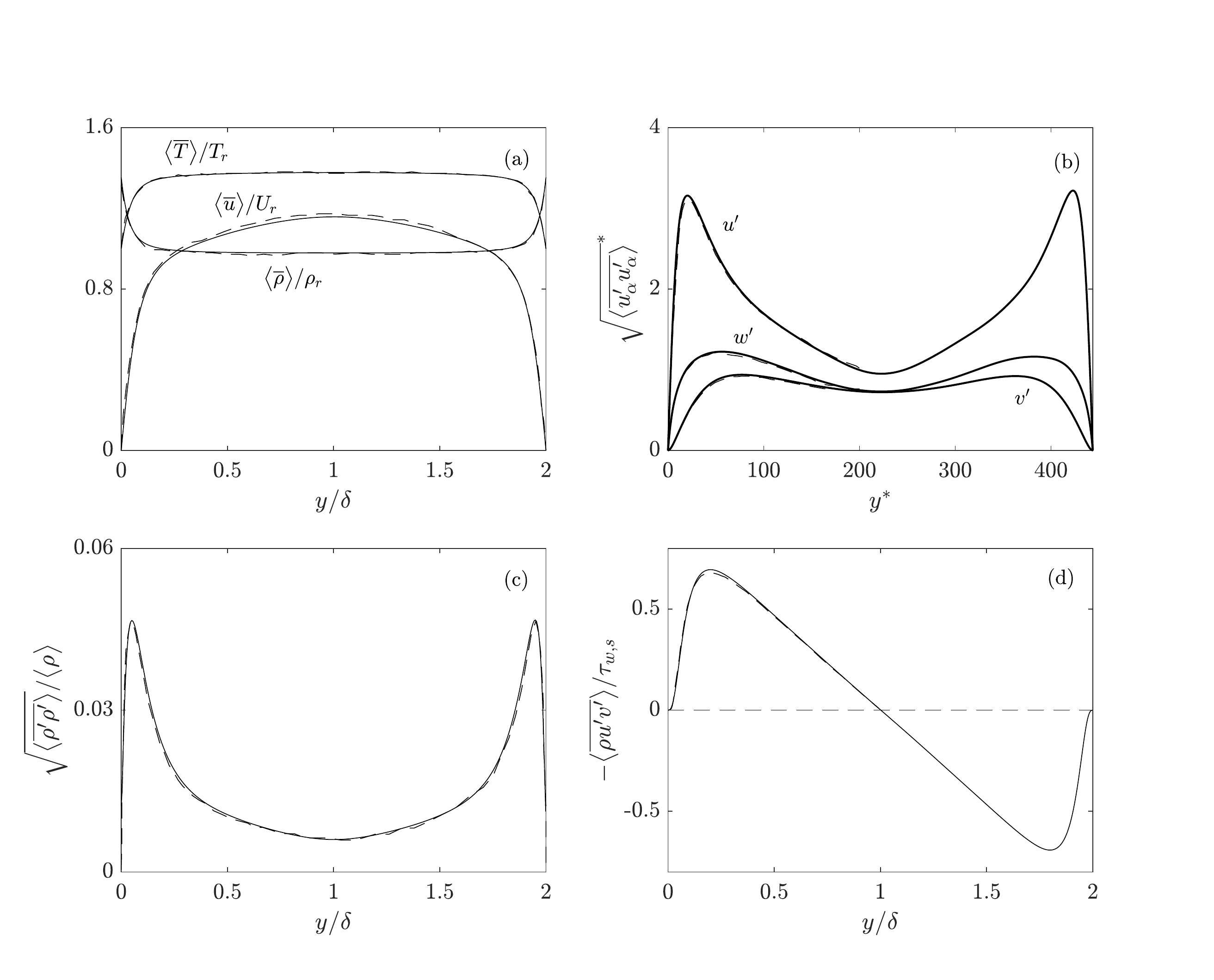}}
  \caption{Profiles of mean and turbulence variables for the smooth-wall flow at $\textup{Re}_b=3000$ and $M=1.5$: $\solid$ present simulation, $\dashed$ \citet{ColemanKM95}. (a) Mean values of temperature, streamwise velocity and density, (b) r.m.s. of turbulent velocities (no summation over Greek indices, $*$ denotes normalization in wall units using $\tau_{w,s}=\mu_r\big<\frac{d\overline{u}}{dy}\big\vert_w\big>$ and $\rho_w$), (c) r.m.s. of density, and (d) Reynolds shear stress.}
\label{fig5:smooth}
\end{figure}

\begin{figure}
   \centerline{\includegraphics[width=1\textwidth,trim={0cm 0cm 0cm 0cm},clip]{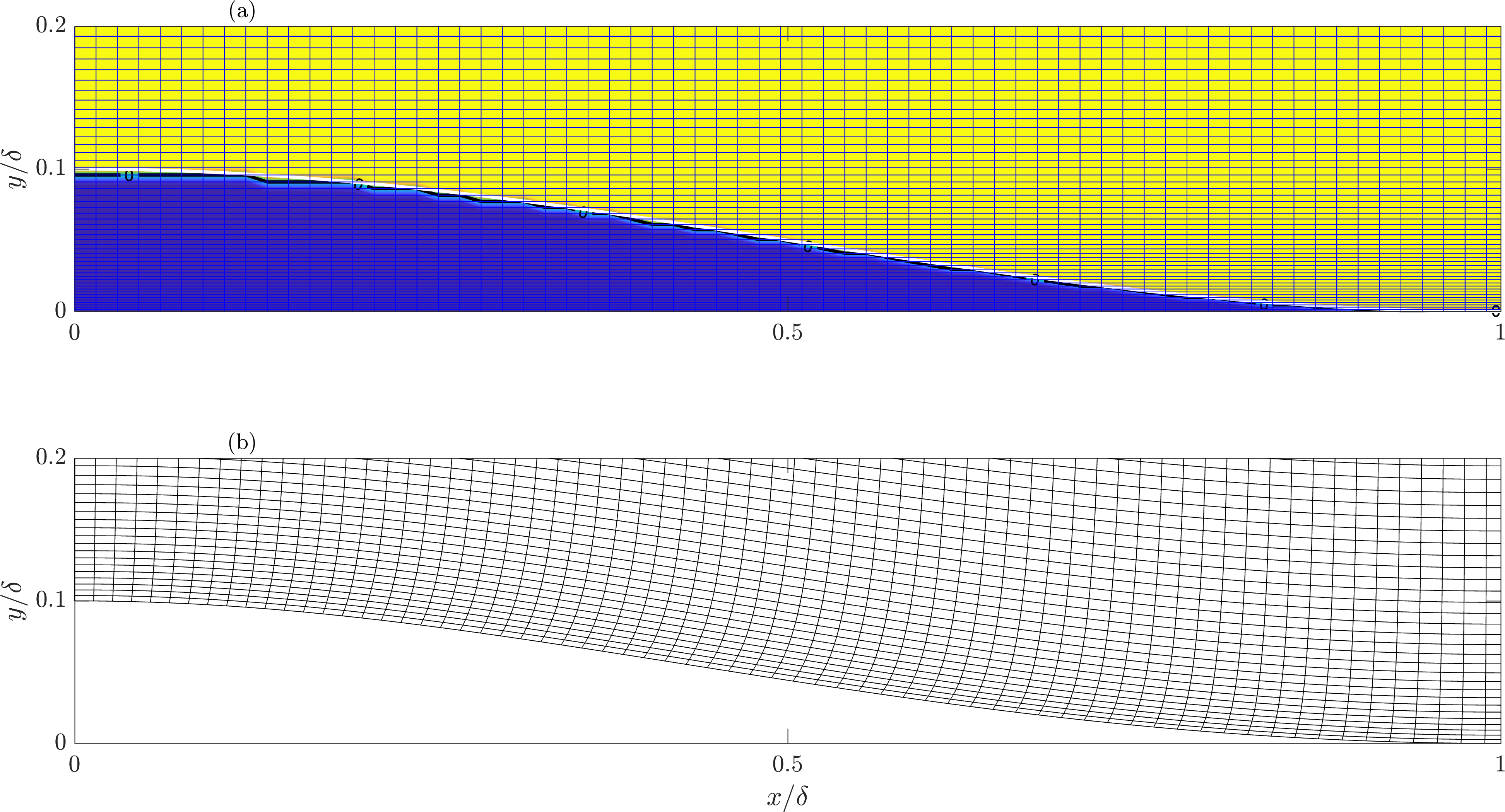}}
  \caption{Contour of level set $\psi$, ranging from -1 (blue) to $+1$ (yellow) used for the IB method (a), and mesh used for the conformal setup (b), both for Case C1. {In subplot (a) the solid white and black lines indicate, respectively, the exact roughness height  as in equation (\ref{eq:hei2d}) and the iso-line of $\psi=0$ obtained from the level-set equation (\ref{eq:psi}); the difference between the two lines is one grid point maximum.}}
\label{fig5:vof_X1}
\end{figure}

\begin{figure}
   \centerline{\includegraphics[width=1\textwidth,trim={0cm 0cm 0cm 0cm},clip]{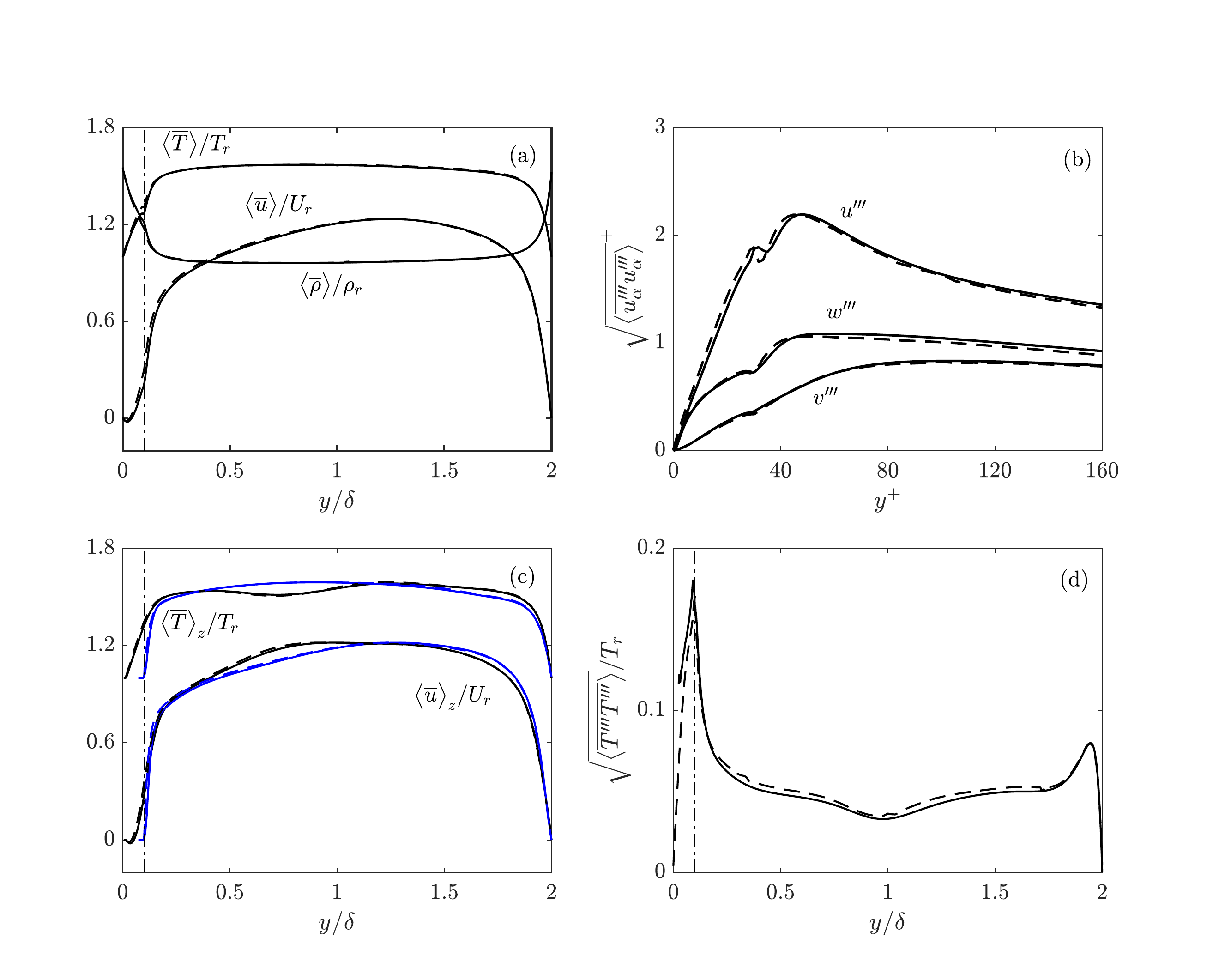}}
  \caption{Mean and turbulence variables for case C1, simulated  using the IB method ($\solid$) and the conformal mesh  ($\dashed$): mean temperature, streamwise velocity and density (a), r.m.s. of velocity components in plus units (roughness side, b), time and spanwise average of velocity and temperature at the roughness crest and valley locations (c, crest in blue and valley in black), and r.m.s. of temperature (d). {In (d), note that  temperature r.m.s. is theoretically zero at the roughness trough  ($y=0$); the intrinsic-averaged value in $y\approx0$ region for the IBM case, however, fluctuates  due to the limited fluid area. This region is removed form the plot.} The vertical dot-dash lines show $y=k_c$. Superscript $+$ denotes normalization in wall units using $u_{\tau,r}$ (tabulated in table \ref{tab:chp5_cf}) and $\rho_r$. }
\label{fig5:validation_X1}
\end{figure}

Periodic BCs are used in the streamwise and spanwise directions. A no-slip iso-thermal wall BC is imposed at both top and bottom walls. The   values of velocity and temperature on both walls  (denoted by subscript $w$)  are $\boldsymbol{u}_w=\boldsymbol{0}$ and $T_w=1$ (the temperature at the wall is used as the reference temperature, i.e $T_r=T_w$). There is no need to impose a BC for density;  { equation (\ref{eq:NS_Compr}a) is solved using third-order accurate one-sided differentiation to update the density values at the boundaries. } This approach is similar to those used in other wall-bounded compressible flow studies \citet[see e.g.][]{ColemanKM95, Tyson13}. The pressure at the boundaries is calculated using the equation of state.

The reference density and velocity used in this work are those of bulk values, defined as $\rho_r\equiv\frac{1}{V_f}\int_{V_f}\overline{\rho}\text{d}v$ and $U_r\equiv\frac{1}{\rho_r V_f}\int_{V_f}\overline{\rho u}\text{d}v$ (where $V_f$ is the fluid occupied volume). The reference length scale  is $\delta$. All these values are set to be 1 here. 
The time-dependent body force $f_1$ in NS equation (\ref{eq:NS_Compr}) is adjusted automatically in each time-step to yield the constant bulk velocity  under the prescribed Reynolds number. 
{ 
Specifically, at each time-step the bulk velocity  $U_r$ is first calculated and  $f_1$ is then determined numerically to compensate for the deviation of $U_r$ from 1.0,   $f_1^{new}=f_1^{old}+\alpha(1-U_r)$, where $\alpha>0$ is taken as a constant of the order of $\rho_r/\text{d}t$ (where $\text{d}t$ is the time-step).
If $U_r<1.0$, $f_1$ increases proportionally to increase $U_r$ in the next time-step, and vice versa. 
This method yields a maximum $|U_r(t)-1.0|$ of $10^{-5}$ for all time-steps after the dynamically steady state is reached.}
The simulations are conducted at $\textup{Re}=3000$ and $M=1.5$, assuming $\textup{Pr}=0.7$ and that the dimensionless viscosity (normalized by its wall value $\mu_r$) and temperature satisfy $\mu=T^{0.7}$.

The respective numbers of grid points in the $x$, $y$ and $z$ directions are $n_x=800$, $n_y=200$ and $n_z=400$. For the present channel size and Reynolds number, the spatial resolution yield $\Delta x^+$, $\Delta y^+_{\max}$ and $\Delta z^+$ less than 3.0, which is sufficiently fine  for DNS. The first 3 grid points in the wall-normal direction are in the $y^+<1.0$ region. The simulations are run in a sufficient amount of simulation time to reach the steady state. Thereafter the statistics are averaged over approximately 20 large eddy turn over time ($\delta/u_{\tau,avg}$, where $u_{\tau,avg}$ is an  average value of the friction velocities on both walls, see table \ref{tab:chp5_cf} for definition).

\subsection{Validation of the numerical method and the  IB method}
\label{sec:valid}

The numerical method is validated by simulating a smooth-wall supersonic turbulent channel flow  at $M=1.5$ and $\textup{Re}=3000$. The same setup was   employed by \citet{ColemanKM95}, which is used here as the benchmark study.

Figure \ref{fig5:smooth}  compares mean and turbulence statistics of the present simulation with those of \citet{ColemanKM95}. The two simulations are in a good agreement for mean velocity, density and temperature, as well as for Reynolds stresses and density variance. This verifies the numerical solver.

To validate the proposed IB method, we  simulated case C1 in two ways: one using the IB method and the other solving the conventional NS equations on a  body-fitted mesh setup. The contour of level set function $\psi$ for the IB method and the mesh of the conformal setup are compared in figure \ref{fig5:vof_X1}. The contour line corresponding to $\psi=0$   represents well the fluid-solid interface. 

Figure \ref{fig5:validation_X1} compares the results obtained using the IB method and those using a body-fitted mesh, in terms of  various mean and turbulence variables, including mean profiles of velocity, temperature and density, Reynolds stresses, and variance of temperature. All  plots show {a good agreement} between the two simulations. 
{ One notice, however, slight differences (about 5\%)   in  the $u'''$ r.m.s. profiles near the crest elevation of roughness (i.e. $y^+= 25$). This is probably due to the different meshes used in the immersed boundary  simulation and the conformal one and  the interpolation scheme used in the conformal case to convert fluid fields to the Cartesian coordinates before the statistics were calculated. 
Overall, these results validate the use of the present IB method to characterize single-point statistics of mean flow and turbulence.  }
See Movie 1 for flow visualizations  for case C1 with the IB method.

\section{Results}\label{sec5:results}

\subsection{Visualizations of the instantaneous and averaged fields}

\begin{figure}
   \centerline{\includegraphics[width=1\textwidth,trim={0cm 0cm 0cm 0cm},clip]{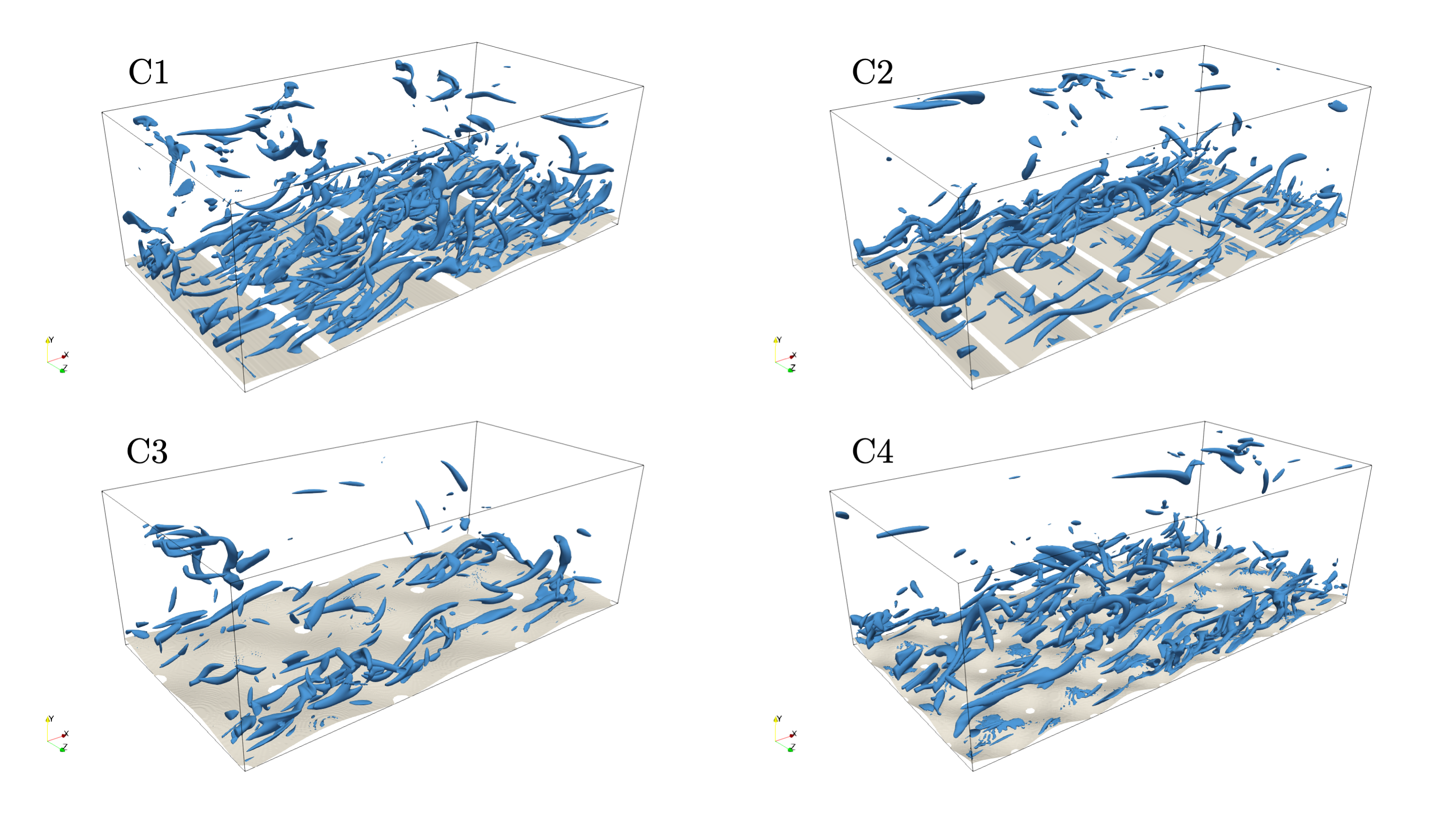}}
  \caption{Isosurfaces of $Q=3$ (in blue, normalized by $U_r$ and $\delta$) for all rough cases. {The gray isosurfaces show the  surface of roughness.}}
\label{fig5:Qcr}
\end{figure}


\begin{figure}
   \centerline{\includegraphics[width=1\textwidth,trim={0cm 0cm 0cm 0cm},clip]{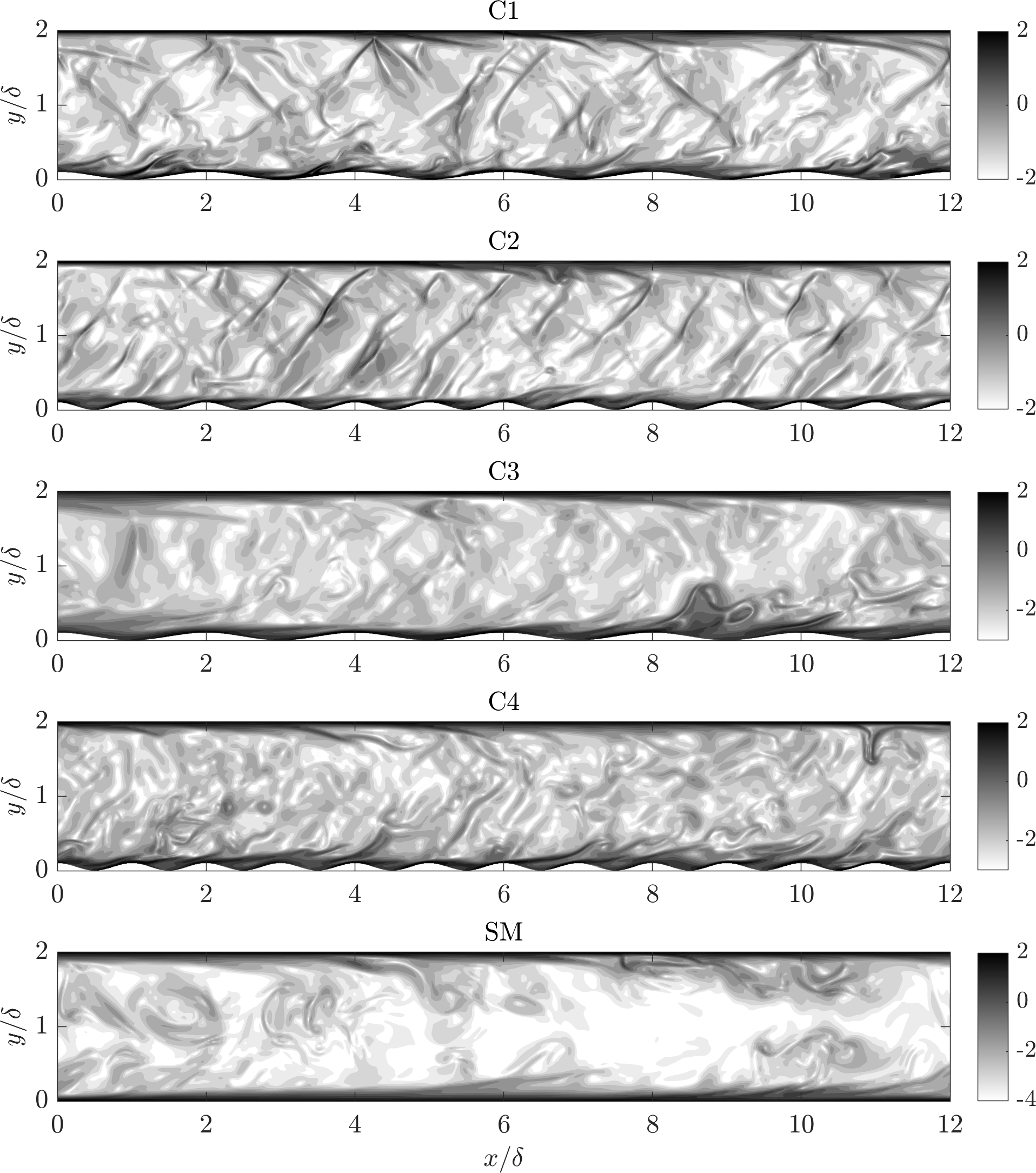}}
  \caption{{Numerical Schlieren images, showing contours of instantaneous $\log{\vert\boldsymbol{\nabla}\rho\vert}$. For a better visualization the  contour ranges are chosen differently for different cases.    $\boldsymbol{\nabla}\rho$ is normalized by $\rho_r$ and $\delta$.} }
\label{fig5:Schil}
\end{figure}


\begin{figure}
   \centerline{\includegraphics[width=1\textwidth,trim={0cm 0cm 0cm 0cm},clip]{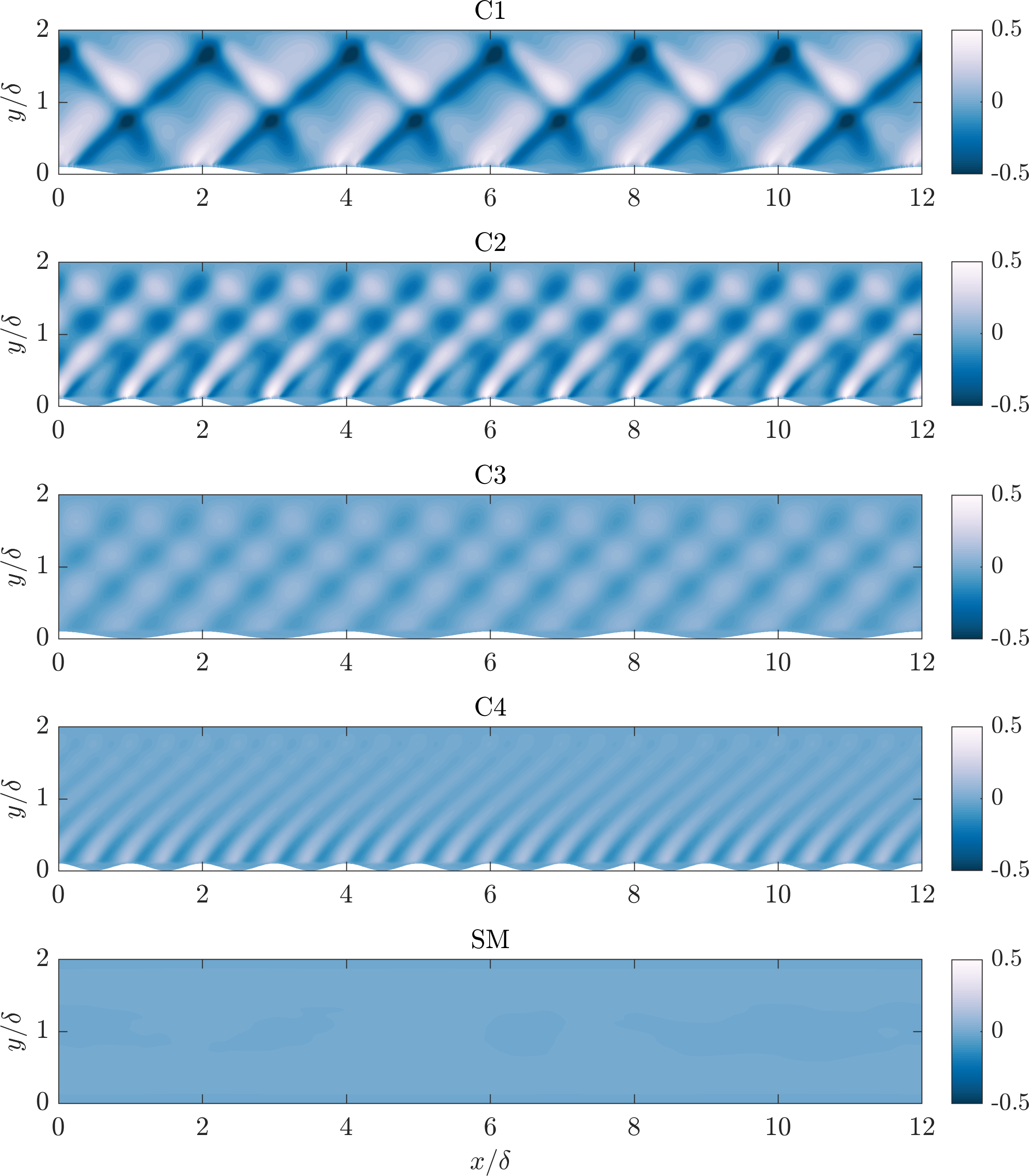}}
  \caption{Contours of $\boldsymbol{\nabla\cdot u}$ averaged in time and spanwise direction. All normalized by $U_r$ and $\delta$. { To calculate the spanwise-averaged values, intrinsic averaging was performed along the spanwise direction at each $(x,y)$ point. An $x$ slice of the corresponding rough surface is shown in each subplot.} }
\label{fig5:div_avg}
\end{figure}

\begin{figure}
   \centerline{\includegraphics[width=1\textwidth,trim={0cm 0cm 0cm 0cm},clip]{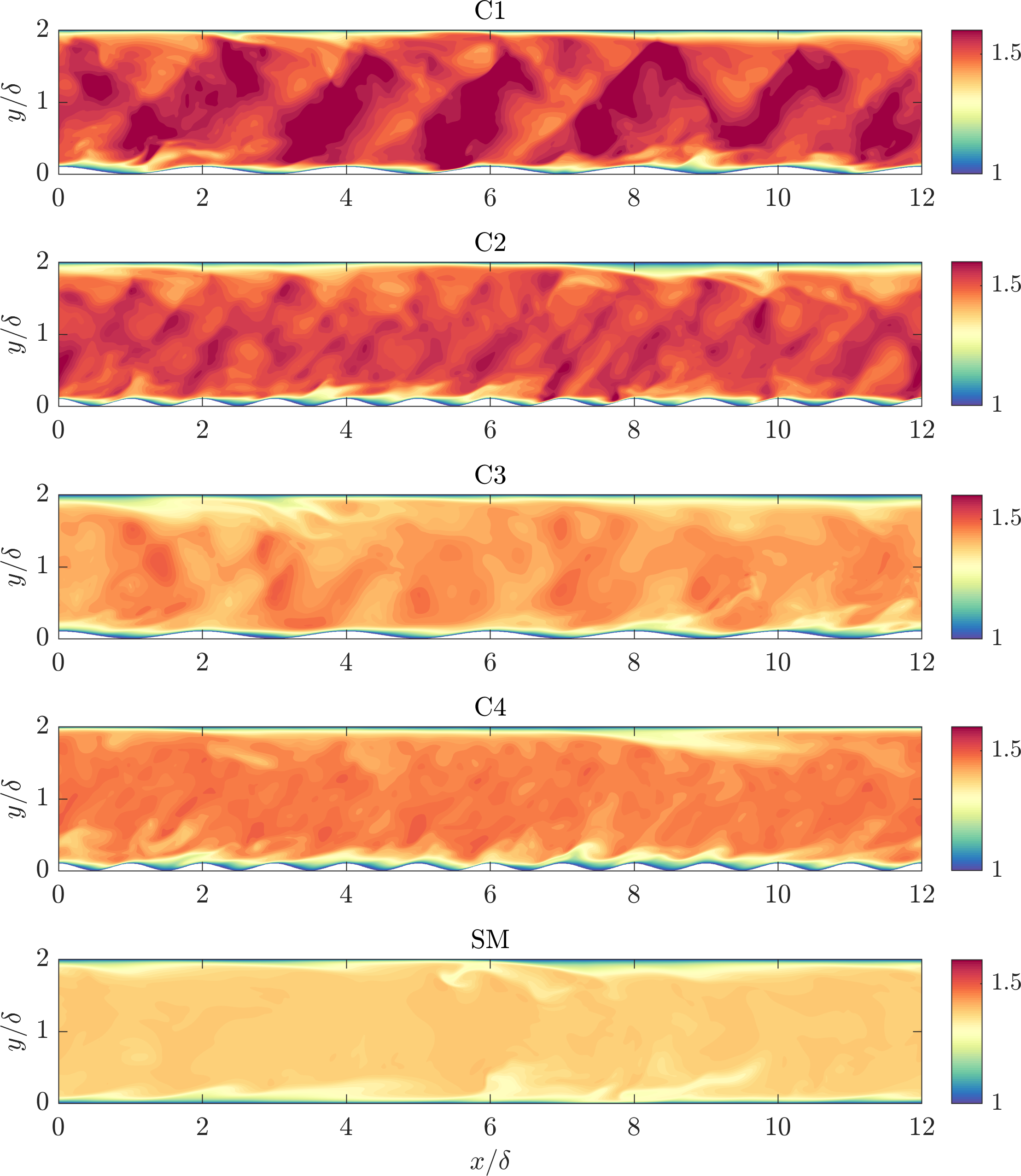}}
  \caption{Contours of instantaneous $T$, normalized by $T_r$.}
\label{fig5:T_cont}
\end{figure}

	\begin{table}
   \begin{center}
     \def~{\hphantom{10}} 
     \begin{tabular}{l c c c c c c c c}
  Case	  & $u_{\tau,s}/U_r$ & $u_{\tau,r}/U_r$  & $u_{\tau,avg}/U_r$ & $\textup{Re}_\tau$	 & $\rho_w$ & $\Delta U^*_{VD}$&$\Delta U^+_{VD}$	& $C_f\times 10^3$	   \\ 
C1	 & 0.0629 & 0.0997 & 0.0834 & 250 & 1.539 & 7.76 &	7.72 & 	13.9	 \\
C2	 & 0.0634 & 0.0912 & 0.0785 & 236 & 1.484 & 6.61 & 6.64	 & 	12.3	\\
C3	 & 0.0642 & 0.0847  & 0.0751 & 225 &1.456 & 5.13 &	5.15 & 	11.3	\\
C4	 & 0.0644 & 0.0899  & 0.0782 & 235  &1.451 &6.25 & 6.3	 & 	12.2	\\
SM	 & 0.0633 & --     & 0.0633 & 190 &1.360 &--&--	 & 	8.0	\\
     \end{tabular}
     \caption{Wall friction comparison. $u_{\tau,s}=\sqrt{\tau_{w,s}/\rho_r}$ and $u_{\tau,r}=\sqrt{\tau_{w,r}/\rho_r}$, where $\tau_{w,s}=-\mu_r\frac{d<\overline{u}>}{dy}\big\vert_{y=2\delta}$ is wall shear stress on the smooth side and
     {$\tau_{w,r}=\frac{1}{A_t}\int_{V_f}\overline{f}_1 \text{d}v -\tau_{w,s}$ is that on the rough side (obtained from momentum balance).} 
     $\textup{Re}_\tau=\rho_r u_{\tau,avg} \delta/\mu_r$, $C_f=2(u_{\tau,avg}/U_r)^2$, $u_{\tau,avg}^2=\big(u_{\tau,s}^2+u_{\tau,r}^2\big)/2$, 
{ $\rho_w$ is the density value at $y=0$, and $\Delta U^*_{VD}$ and $\Delta U^+_{VD}$ are the roughness functions associated with the Van Driest transformed velocities.}
}
     \label{tab:chp5_cf}
     \end{center}
     \end{table}

Instantaneous vortical structures are visualized using iso-surfaces of $Q$-criterion \citep{Chong90} in figure \ref{fig5:Qcr}. 
Modifications of the near wall turbulence  on the rough-wall side are noticeable. { The main difference between the effects of different roughness geometries  is the shock patterns shown by  the instantaneous numerical Schlieren images (figure~\ref{fig5:Schil}).} These patterns are also persistent in time as shown by time- and spanwise-averaged  $\boldsymbol{\nabla\cdot u}$ (figure~\ref{fig5:div_avg}). Both figures  show that 2D surfaces (cases C1 and C2) induce strong shocks  that reach  the upper wall and are reflected back to the domain after impingement. The shock patterns exhibit the same wavelength of the roughness geometries, and   influence the flow properties in the whole channel. This is obvious in the contours of instantaneous temperature fields in figure \ref{fig5:T_cont}, where temperature periodically changes in the compression and expansion regions associated with roughness geometries in C1 and C2. For 3D cases  the embedded shocks  are weaker and, consequently, replaced by the small-scale shocklets.


\subsection{Mean and turbulence variables}

First, the values of frictional velocities on the smooth and rough sides as well as the frictional Reynolds number $\textup{Re}_\tau$ and drag coefficient $C_f$ are tabulated in table \ref{tab:chp5_cf} for all cases. On the rough-wall side, the wall friction include both viscous and pressure drag components.  { Due to  differences in wall friction generated by   roughnesses of different geometries,  $\textup{Re}_\tau$ varies from 190 to 250. Yet, the flows are all  low-Reynolds-number ones; the differences in  shock features and flow statistics (discussed thoroughly below) are thus  likely a result of the change in roughness geometry, instead of the change in friction Reynolds number.}
 
The comparison shows that, as expected, the  wall friction on the rough-wall side is higher than that on the smooth wall side for all cases. 
Overall, a 2D roughness generates  higher friction than a 3D one of the same height. This is consistent with observations in incompressible flows \citep[e.g. by][]{VolinoVSF11} that 2D   roughness affects turbulence more strongly due to the larger length scale (in $z$) that is imparted to the flow.
In addition, results show that  for 3D roughnesses  a higher friction is obtained for a shorter wavelength (or higher roughness slope), which is also consistent with observations in incompressible flows~\citep{NapoliNAD08a}. 
{ It is surprising, however, that between the two 2D rough surfaces the one with a higher slope (C2) yields a lower wall friction.
As will be shown later in section~\ref{sec:budgets}, this appears to be a result of stronger turbulent mixing 
above the rough surface in C1 than in C2, due to regions with more intense compression (those with strong negative values of $\boldsymbol{\nabla}\cdot\boldsymbol{u}$, figure~\ref{fig5:div_avg}) in which strong turbulent-kinetic-energy (TKE) production and redistribution take place.
These observations indicate that, in fully developed supersonic rough-wall flows, the dependency of the wall friction on  roughness geometry is more complex than in incompressible flows, due to compressibility effects.  Future systematic studies with a wide range of different rough surfaces  are needed to detail the dependences of shocks  on  roughness height and geometry, and  analyses of near-wall momentum balance are needed to further understand changes of the flow.
}

\begin{figure}
   \centerline{\includegraphics[width=1\textwidth,trim={0cm 0cm 0cm 0cm},clip]{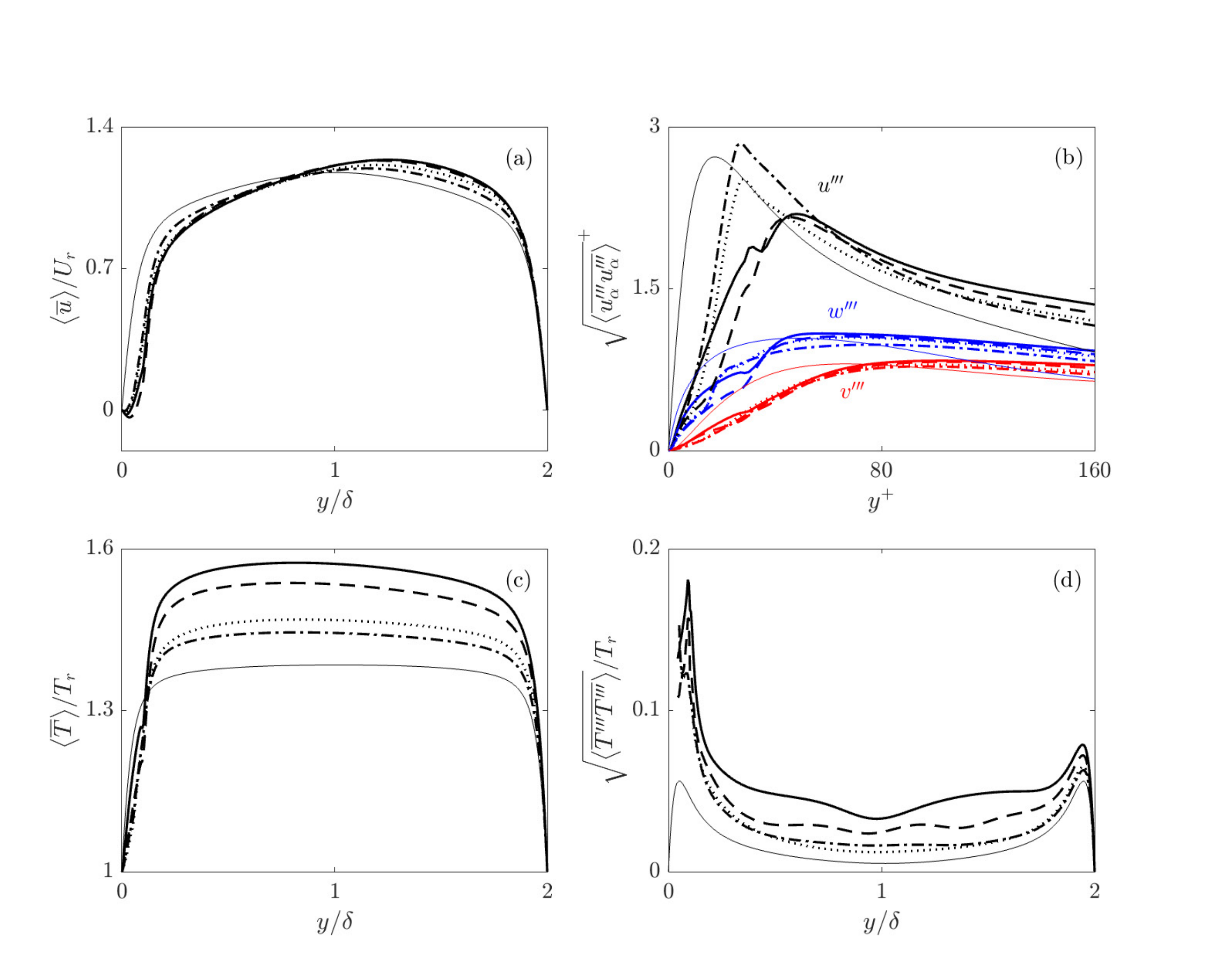}}
  \caption{Mean and turbulence variables  for cases C1 ($\solid$), C2 ($\dashed$),  C3 ($\chndot$), C4 ($\dotted$) {and SM ($\solidd$)}: profiles of the double-averaged streamwise velocity (a), r.m.s. of velocities  in plus units (roughness side, b), double-averaged  temperature (c), and r.m.s. of temperature (d). In (d), note that  temperature r.m.s. is theoretically zero at the roughness trough  ($y=0$); the intrinsic-averaged value in $y\approx0$ region, however, fluctuates  due to the limited fluid area. This region is removed form the plot.}
\label{fig5:mean_turb}
\end{figure}

Figure \ref{fig5:mean_turb} compares profiles of the mean and turbulence quantities between different   cases. The mean streamwise velocity (figure \ref{fig5:mean_turb}a) and density (not shown) are both weakly dependent on the roughness geometry across the channel, except for the region near the rough wall. This is because the normalization  using the bulk values ($U_r$ and $\rho_r$)   absorbs major differences in the velocity and density profiles in the bulk part of the channel. 
The mean temperature values (figure \ref{fig5:mean_turb}c), on the other hand, differ across the channel for different roughness topographies. The mean temperature is higher for {2D roughness cases} (C1 and C2) compared to the 3D ones (C3 and C4) and the smooth case (SM). Here, the temperature  is normalized by the wall value,  $T_w$, which does not  absorb the differences in the core region. { It is  established \citep[][chapter 3]{Anderson90} that shock waves result in entropy generation, because of  strong viscous effects and thermal conduction   in large gradients regions.    }
The stronger shocks  in the 2D roughness cases involve more entropy in the domain than in the 3D cases. As a result, the irreversible heat generation is more intense for these cases, leading to higher temperature values.

{The r.m.s. of the three $u'''_i$ fluctuation components   are plotted in  figure \ref{fig5:mean_turb}(b) in wall units}.
For rough cases, it shows that roughness effects are { mostly} confined to a near-wall  region; outside this region  { the differences between profiles for different rough surfaces are smaller} for all velocity components. This is similar to the concept of roughness sublayer { (defined as the near-wall layer where turbulence statistics in wall units vary  with the wall condition \citep{FlackSC07}) } in an incompressible turbulent flow bounded by rough wall. Near the wall, the $v'''$ and $w'''$ components are similar among all cases, whereas  the $u'''$ components in 3D cases display a peak closer to the wall than their 2D counterparts. Similar phenomena were observed for incompressible flow; it was explained as a result of a thicker roughness sublayer over a 2D roughness \citep{VolinoVSF11}, leading to a peak farther from the smooth-wall peak elevation at  $y^+\approx 15$. 
{
The fact that the turbulence intensities in wall units do not collapse perfectly in the outer layer (i.e. the region above the roughness sublayer) among all cases indicates that the wall similarity  \citep[or ``outer layer similarity'',][]{SchultzF07} of \citet{Townsend76} does not apply. The wall similarity hypothesis (primarily describing incompressible flows) states that, at high Reynolds number and with very small roughness compared to $\delta$, turbulent statistics  outside the roughness sublayer are independent of wall roughness, except for its scaling on the friction velocity.
Given the relatively low Reynolds number and large roughness ($k_c/\delta = 0.1$) in the present cases, exact wall similarity is not expected. In addition, in current supersonic flows roughness is shown to directly affect outer layer turbulence through its effect on shocks that extend to the core region of the channel flow (figures~\ref{fig5:Schil} and \ref{fig5:div_avg}).

}

The profiles of  temperature r.m.s. in figure \ref{fig5:mean_turb}(d) show that the intensity of temperature fluctuations far from the wall depends strongly on the roughness geometry. For {2D rough surfaces}, the variations of curve shape in the bulk of the channel are associated with the shock patterns in the domain. Temperature varies significantly near the locations where the shock waves coincide and form  nodes of  shock diamonds (i.e. the nodes   away from walls). These shock diamonds are also visible in figures \ref{fig5:T_cont}, \ref{fig5:Schil} and \ref{fig5:div_avg} (C1 and C2). For 3D cases the shock diamonds are weak or nonexistent. Therefore, the curves of temperature r.m.s. in 3D cases are smooth in the core region. 
      
\begin{figure}
   \centerline{\includegraphics[width=1\textwidth,trim={0cm 0cm 0cm 0cm},clip]{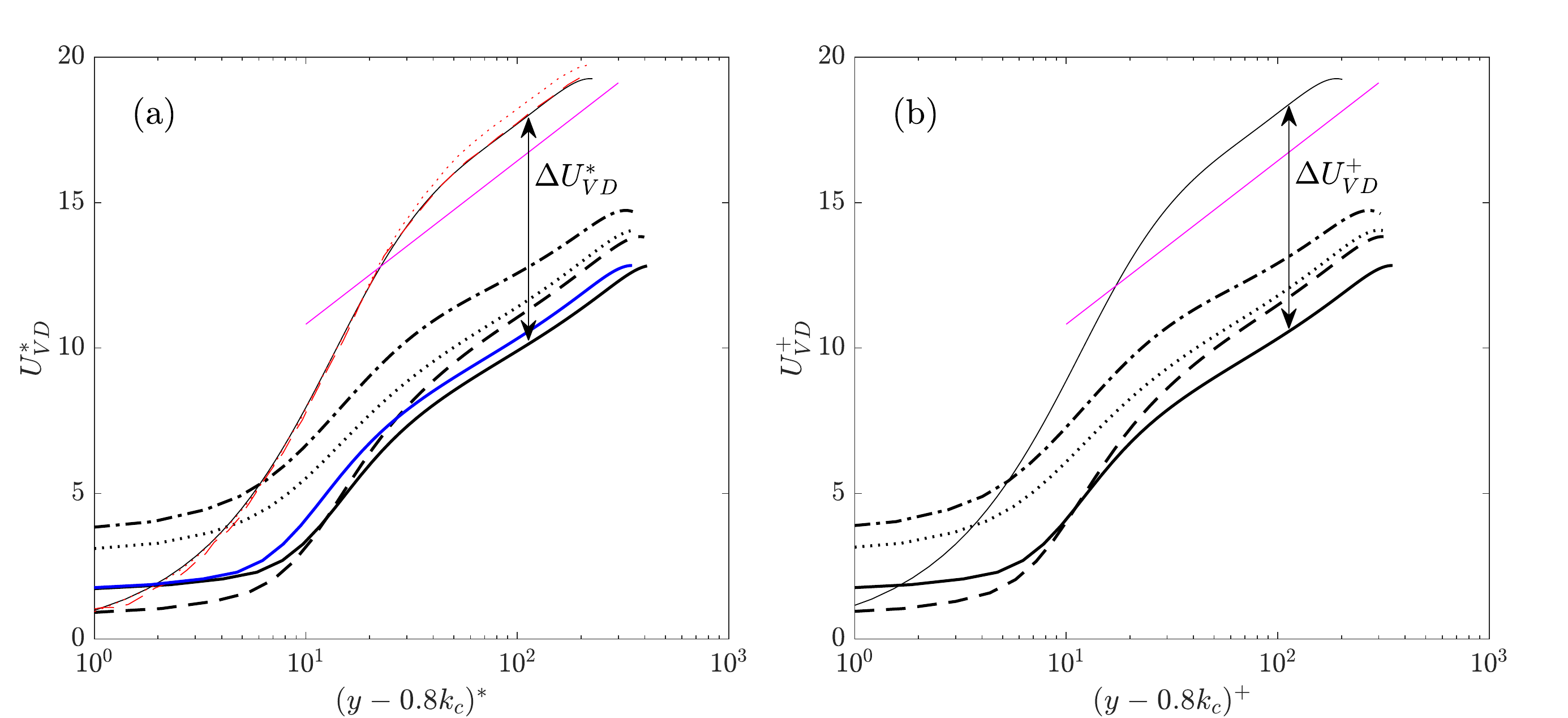}}
  \caption{{
Law of the wall. Profiles of mean velocities transformed using (a) original Van Driest transformation (equation~\ref{e:VD1})  and (b) a modified Van Driest transformation (equation~\ref{e:VD2}). Cases C1 ({\color{black}$\solid$}), C2 ({\color{black}$\dashed$}),  C3 ({\color{black}$\chndot$}), C4 ({\color{black}$\dotted$}) and SM ({\color{black}$\solidd$}). 
Solid magenta lines ({\color{magenta}$\solid$}) shows slope of $1/\kappa$, where $\kappa=0.41$ is the von Karman constant. In subplot (a) results of \citet{ColemanKM95} ({$\dashed$}) and \citet{Foysi04} ({$\dotted$}) for smooth-wall flows are provided for comparison, and the blue solid line ({$\solid$}) is same as the  C1 profile in subplot (b). $\Delta U^*_{VD}$ and $\Delta U^+_{VD}$ are roughness functions for case C1. } }
\label{fig5:uvd}
\end{figure}      
      
{Figure \ref{fig5:uvd} compares the mean velocity profiles in inner units.
The law of the wall for mean velocity scaled in this way refers to the universal   logarithmic profile in regions between $50\delta_\nu\lesssim y\lesssim0.2\delta$  (where $\delta_\nu$  is the viscous length scale) on a smooth wall. On a rough wall, incompressible flow studies showed that the logarithmic profile still exists, with its lower extent shifted to the top of roughness sublayer.
Both roughness and compressibility were found to influence the law of the wall through their effects on the inner units. 
Specifically, roughness shifts the logarithmic profiles downward for an amount  $\Delta U^+$ (called roughness function) with respect to a smooth-wall flow \citep{Nikuradse33e}. This has been observed  for a wide range of roughness topographies  \citep[][to name a few]{RaupachAR91,SchultzF07,Leonardi07,ForooghiFSMJF17a, BusseBTS17,Womack22}.

For compressible flows, finding appropriate inner velocity, length, density and viscosity scales that result in universal law of the wall  is an active subject of research for smooth-wall flows \citep[][among many others]{Morkovin62, Volpiani20}. The complexities stem from significant variations in density, viscosity and heat transfer across the boundary layer that need to be accounted for. 
Here we plot density-transformed mean velocity profiles, introduced by \citet{Van-Driest51}. The approach has been shown to collapse  mean velocity profiles for smooth wall flows at different Mach numbers \citep{Guarini2000,Lagha11,Trettel16}. The original Van Driest transformation reads as 
\begin{equation}
\label{e:VD1}
U^*_{VD}=\int_0^{\langle\overline{u}\rangle^*}\bigg(\frac{\langle\overline{\rho}\rangle}{\rho_w}\bigg)^{1/2}\text{d}\langle\overline{u}\rangle^*,
\end{equation}
where superscript $*$ denotes normalization using $\tau_{w,r}$, $\rho_w=\langle\overline{\rho}\rangle\vert_{y=0}$ and $\mu_r$. The results are plotted in figure \ref{fig5:uvd}(a). The profiles of \citet{ColemanKM95}  and \citet{Foysi04} for   smooth-wall flows ($M=1.5$ and $\text{Re}=3000$) are also compared. The present SM profile matches very well with the reference data. We also employ a modified Van Driest transformation, where all density scales are normalized by  $\rho_r$ instead of $\rho_w$. The modified Van Driest transformation is
\begin{equation}
\label{e:VD2}
U^+_{VD}=\int_0^{\langle\overline{u}\rangle^+}\bigg(\frac{\langle\overline{\rho}\rangle}{\rho_r}\bigg)^{1/2}\text{d}\langle\overline{u}\rangle^+,
\end{equation}
where superscript $+$ denotes normalization using $\tau_{w,r}$, $\rho_r$ and $\mu_r$. The results are plotted in figure \ref{fig5:uvd}(b). Since both $\rho_w$ and $\rho_r$ are in the order of unity,   $U^*_{VD}$  and $U^+_{VD}$ are  not  significantly different (comparing the solid blue and black lines in figure \ref{fig5:uvd}a for case C1). A constant displacement height $d=0.8k_c$ is chosen  for the rough cases in figure \ref{fig5:uvd}.
All rough-wall profiles in figure \ref{fig5:uvd} show a downward shift ($\Delta U_{VD}$) with respect to the smooth wall due to higher wall friction, similar to incompressible flows. The magnitudes of  $\Delta U^*_{VD}$ and $\Delta U^+_{VD}$  are measured at $(y-0.8k_c)^*=100$ and $(y-0.8k_c)^+=100$, respectively, and tabulated in table~\ref{tab:chp5_cf}. The two different transformations give virtually the same roughness functions, which  are larger for 2D rough cases than 3D ones and display the same comparison as that of   $C_f$ among all cases (table \ref{tab:chp5_cf}). 
These observations suggest that the discussions in the literature concerning  equilibrium incompressible  rough-wall drag laws  may be  extendable to equilibrium supersonic  rough-wall flows, when the Van Driest types of transformation are employed, as long as the Mach number is not too high. The latter may be necessary as   essential dynamics of turbulence in equilibrium compressible flows are expected to remain similar to their incompressible counterparts, if the Mach number is not high enough to yield  prevailing  compressibility effects  \citep{Morkovin62}. Also, one notices in figure \ref{fig5:uvd} that $1/\kappa$ (where $\kappa\approx0.41$ being the von Karman constant) is still a good approximation  for the slopes of both $U^*_{VD}$ and $U^+_{VD}$ profiles for the present rough cases, though with minor noticeable variations.

}      
\subsection{Budgets of the Reynolds stresses}\label{sec:budgets}

The transport equation for various components of the Reynolds stress tensor in a compressible flow reads as \citep{Vyas19} 
\begin{equation}\label{eq:budg}
\begin{aligned}
    \frac{\partial}{\partial t}(\overline{\rho u_i''u_j''}) =\ & \mathcal{C}_{ij}+\mathcal{P}_{ij}
    +\mathcal{D}_{ij}^{M}+\mathcal{D}_{ij}^{T}+\\
    & \mathcal{D}_{ij}^{P}+{\Pi}_{ij}+\mathcal{\epsilon}_{ij}+\mathcal{M}_{ij},
\end{aligned}
\end{equation}
where $i$, $j$ = \{1, 2, 3\} and $\mathcal{C}$, $\mathcal{P}$, $\mathcal{D}^M$, $\mathcal{D}^T$, $\mathcal{D}^P$, $\Pi$, $\epsilon$ and $\mathcal{M}$, are, respectively, mean convection, production, molecular diffusion, turbulent diffusion, pressure diffusion, pressure-strain, dissipation, and turbulent mass flux terms, and are defined as


\begin{equation}\label{eq:budg-terms}
\begin{aligned}
    \mathcal{C}_{ij}= &\ -\frac{\partial}{\partial x_k}(\overline{\rho u''_iu''_j}\widetilde{u}_k),\\
     \mathcal{P}_{ij}= &\ -\overline{\rho u''_iu''_k}\frac{\partial\widetilde{u}_j}{\partial x_k} -\overline{\rho u''_ju''_k}\frac{\partial\widetilde{u}_i}{\partial x_k},\\
     \mathcal{D}_{ij}^M= &\ \frac{\partial}{\partial x_k}(\overline{u''_i \tau_{kj}}+\overline{u''_j \tau_{ki}}),\\
     \mathcal{D}_{ij}^T= &\ -\frac{\partial}{\partial x_k}(\overline{\rho u''_i u''_j u''_k}),\\
     \mathcal{D}_{ij}^P= &\ -\frac{\partial}{\partial x_k}(\overline{p'u''_i}\delta_{jk}+\overline{p'u''_j}\delta_{ik}),\\
     \Pi_{ij}= &\ \overline{p'\bigg(\frac{\partial u''_i}{\partial x_j}+\frac{\partial u''_j}{\partial x_i}\bigg)},\\
     \epsilon_{ij}= &\ - \overline{\tau_{ki}\frac{\partial u''_j}{\partial x_k}} - \overline{\tau_{kj}\frac{\partial u''_i}{\partial x_k}},\\
     \mathcal{M}_{ij}= &\ \overline{u''_i}\bigg(\frac{\partial\overline{\tau}_{kj}}{\partial x_k}-\frac{\partial \overline{p}}{\partial x_j}\bigg)
     +\overline{u''_j}\bigg(\frac{\partial\overline{\tau}_{ki}}{\partial x_k}-\frac{\partial \overline{p}}{\partial x_i}\bigg).
\end{aligned}
\end{equation}

 \begin{figure}
   \centerline{\includegraphics[width=1\textwidth]{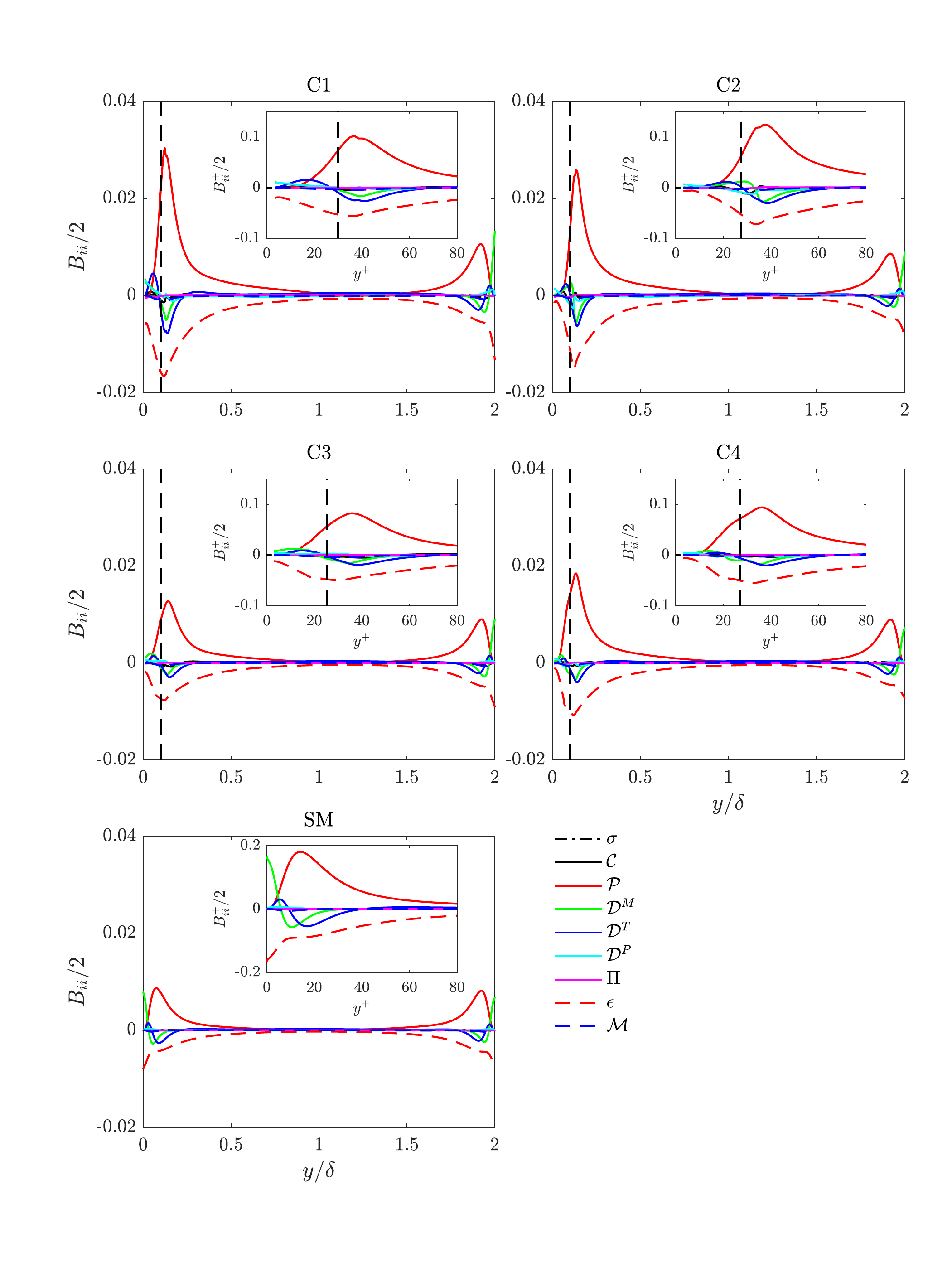}}
  \caption{Budget balances of TKE. All terms are double-averaged in time and in the $x$-$z$ plane. {They are normalized by the outer  units $\rho_r$, $U_r$ and $\delta$ for the external subplots, and wall ($+$) units $\rho_r$, $u_{\tau,r}$ and $\mu_r$ in the insets.} The vertical dash lines show $y=k_c$.}
\label{fig5:TKE}
\end{figure}

\begin{figure}
\centerline{\includegraphics[width=1\textwidth]{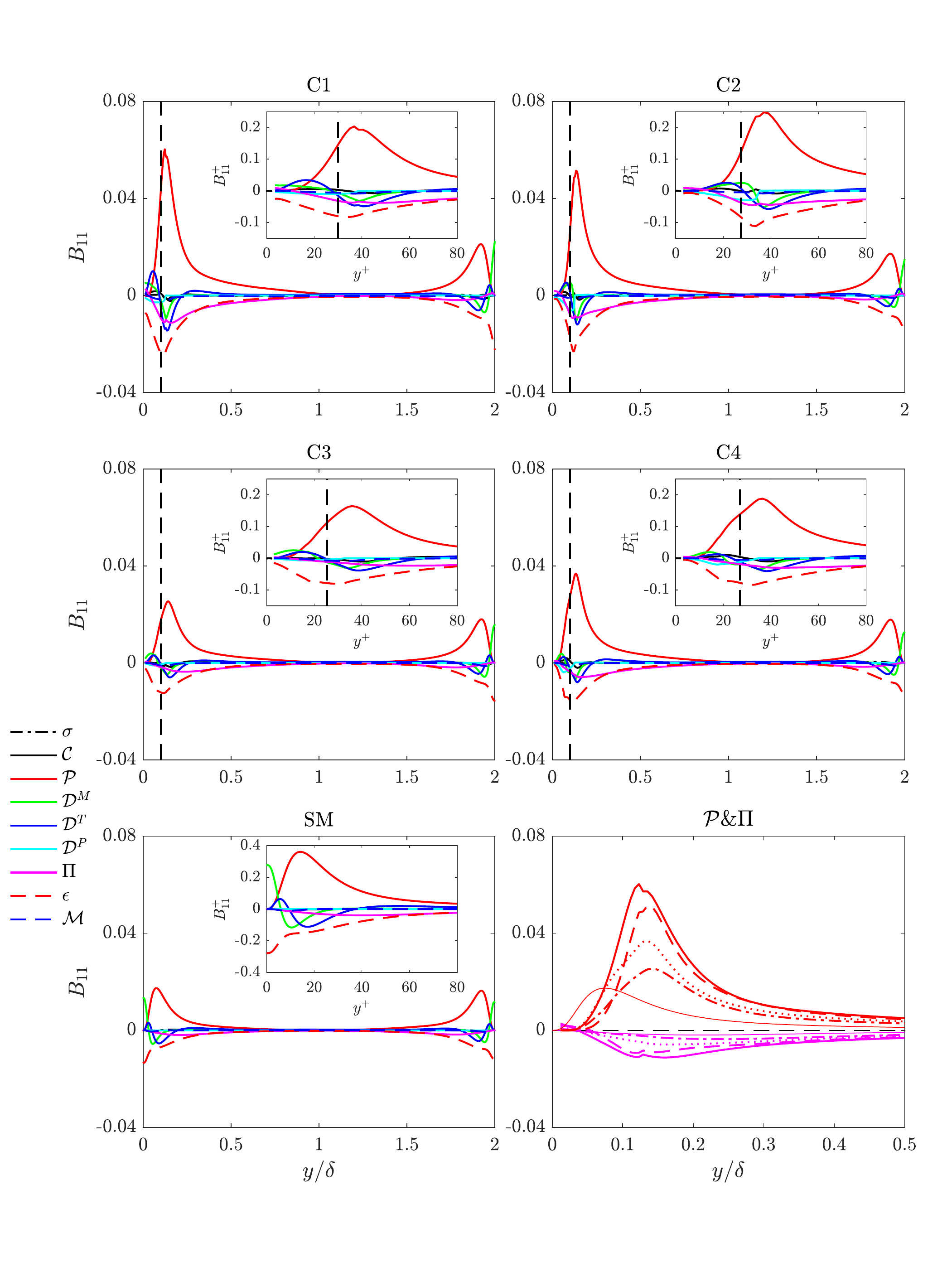}}
  \caption{Budget balances of B11. All terms are double-averaged in time and the $x$-$z$ plane. {They are normalized by the outer  units $\rho_r$, $U_r$ and $\delta$ for the external subplots, and wall ($+$) units $\rho_r$, $u_{\tau,r}$ and $\mu_r$ in the insets. Subplot $\mathcal{P}\&\Pi$ compares the production and  pressure-strain terms of all cases: C1 ($\solid$), C2 ($\dashed$),  C3 ($\chndot$), C4 ($\dotted$) and SM ($\solidd$).} The vertical dash lines show $y=k_c$.}
\label{fig5:B11}
\end{figure}

The budget terms are calculated for all non-zero components of the Reynolds stress tensor and for TKE. The budget balance of the transport equation of $\langle \overline{\rho u''_iu''_j}\rangle$ is denoted as $B_{ij}$. Figures \ref{fig5:TKE} and \ref{fig5:B11} show   wall-normal profiles of the spatial-averaged budget terms of TKE and   $\langle \overline{\rho u''u''}\rangle$, respectively. { Both normalizations by  the reference units ($\rho_r$, $U_r$ and $\delta$) and by the  wall units   ($\rho_r$, $u_{\tau,r}$ at bottom wall and $\mu_r$) are used. 
} The residual of the calculated budget balance, $\sigma$, is presented; it  is less than 1\% of the maximum value of the shear production $\mathcal{P}$ in all cases. { This suggests that the budget terms are calculated correctly and  that the numerical dissipation (as a result of both the solver's flux-splitting procedure and the IBM)  is small for estimation of the budget terms.}

\begin{figure}
   \centerline{\includegraphics[width=1\textwidth,trim={0cm 0cm 0cm 0cm},clip]{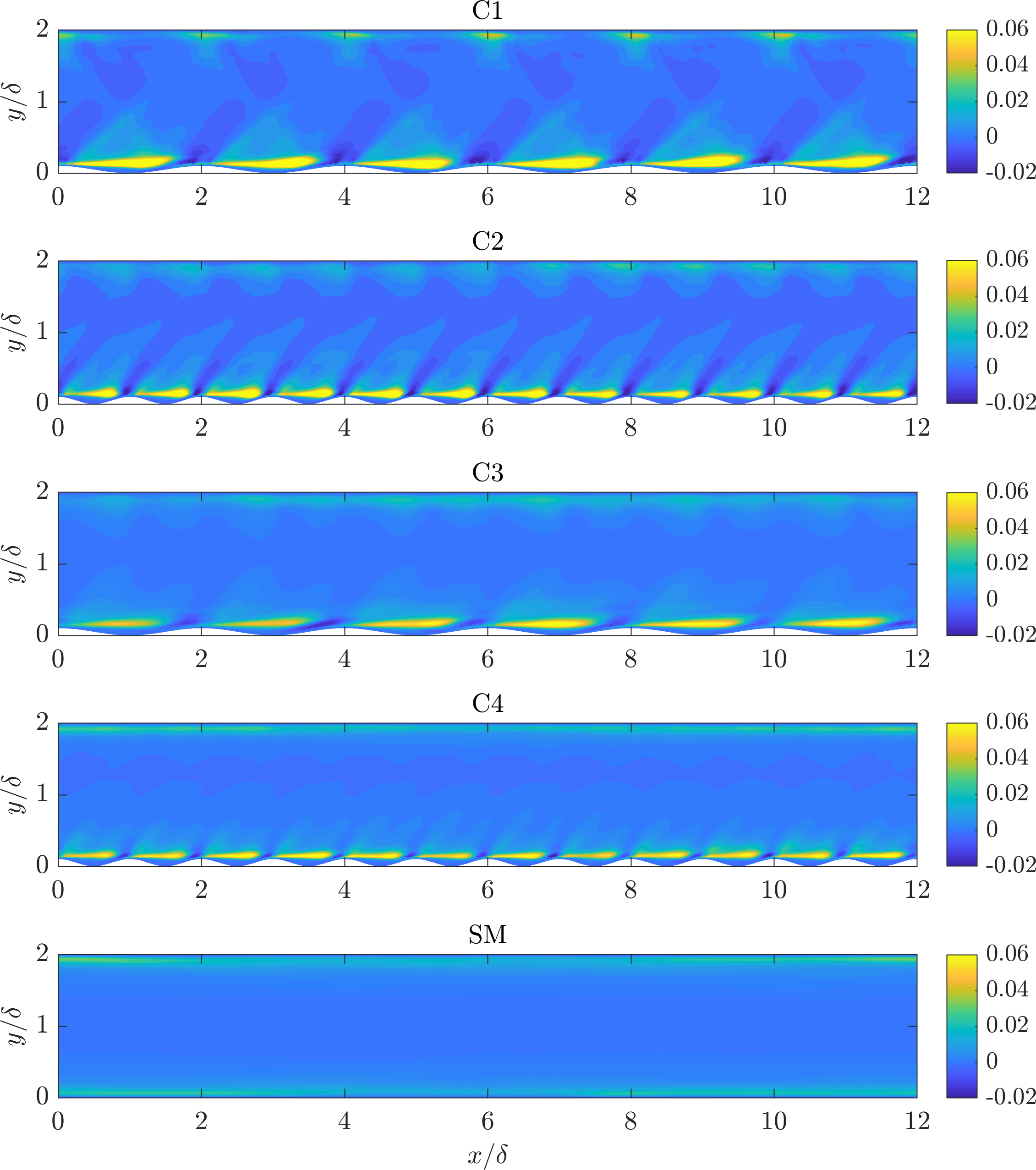}}
  \caption{Contours of ${\mathcal{P}}_{11}$ normalized using $\rho_r$, $U_r$ and $\delta$. An $x$ slice of the corresponding rough surface is shown in each subplot.}
\label{fig5:Pxx_cont}
\end{figure}


Comparing the smooth- and rough-wall cases, the main differences are seen near the  wall. Specifically, molecular diffusion and viscous dissipation are non-zero on the smooth wall, whereas they are zero on the bottom of the rough wall attributed to a quiet region without turbulence at the root of the roughness elements.  
Overall, both production and pressure-strain term on a rough wall peak at elevations near the roughness crest, independent of  the smooth-wall peak elevations.

Among the rough cases, the magnitudes of the budget terms normalized by the reference values are shown to be modified by the roughness topography, with the 2D surfaces producing higher magnitudes 
than the 3D ones. { Two important terms, $\mathcal{P}_{11}$ and $\Pi_{11}$ are compared in figure \ref{fig5:B11}  among all cases.
The comparison among the magnitudes of both $\mathcal{P}_{11}$ and $\Pi_{11}$ displays the same trend as those of the temperature profiles (figure \ref{fig5:mean_turb}c) and $C_f$ (table \ref{tab:chp5_cf}), suggesting that an enhancement of turbulence processes augments  temperature and hydrodynamic drag. }
For further explanations, the contours of $\mathcal{P}_{11}$ in an $(x,y)$ plane are shown in figure \ref{fig5:Pxx_cont} to compare the spatial distribution of this term. 
{
It shows that 2D roughness elements  lead to stronger turbulence production  downstream of each roughness peak than the 3D roughnesses. This is probably because the 2D roughnesses induce   organized recirculation regions that are aligned in $z$ and stronger shear layers around the recirculation regions. 
}
In addition,   turbulence production above a 2D roughness is enhanced by the strong mutual interaction between shock waves. In {2D roughness cases}  the regions where two oblique shock waves impinge together (figure~\ref{fig5:div_avg}) are associated with enhanced turbulence production, whether it is on the rough- or smooth-wall side.
 
The effect of shocks on turbulence is an important phenomenon and represents a fundamental difference between supersonic and subsonic turbulent flows over rough walls -- for incompressible flows most of the roughness effects are confined to near wall regions and the outer layer is expected to be independent of the wall condition (except for the scaling of outer-layer statistics on the friction velocity), also known as  outer layer similarity \citep{Townsend76,RaupachAR91,SchultzF07}. 
However, for the supersonic cases herein,  the effects of wall roughness propagate across the channel and modify turbulence production in the upper wall region via the generated oblique shocks. The same process occurs on the rough-wall side, where the reflected shocks  from the smooth-wall side  impinge back to the rough wall  and enhance the turbulence production in these regions. In other words, turbulence processes on both walls  depend on the interaction of shocks, which  are themselves dependent on the roughness topography. This  indicates that   outer layer similarity  does not apply to such supersonic channel flows, {at least for the Reynolds number, Mach number and wall roughness in the present simulations}. The far-reaching effect of surface details may be of potential use in flow and turbulence control.




\subsection{Conditional analysis}
%




In this section, contributions from  regions { of either expansion, compression or solenoidal flow} to the overall TKE production is analyzed and compared among all cases. The velocity divergence  ($\boldsymbol{\nabla}\cdot\boldsymbol{u}$) is used as a measure of compressibility. Large magnitudes of $\boldsymbol{\nabla}\cdot\boldsymbol{u}$ correspond to regions of strong compression or expansion immediately before and after a shock wave. 


\begin{figure}
   \centerline{\includegraphics[width=.95\textwidth,trim={0 0cm 0 0cm},clip]{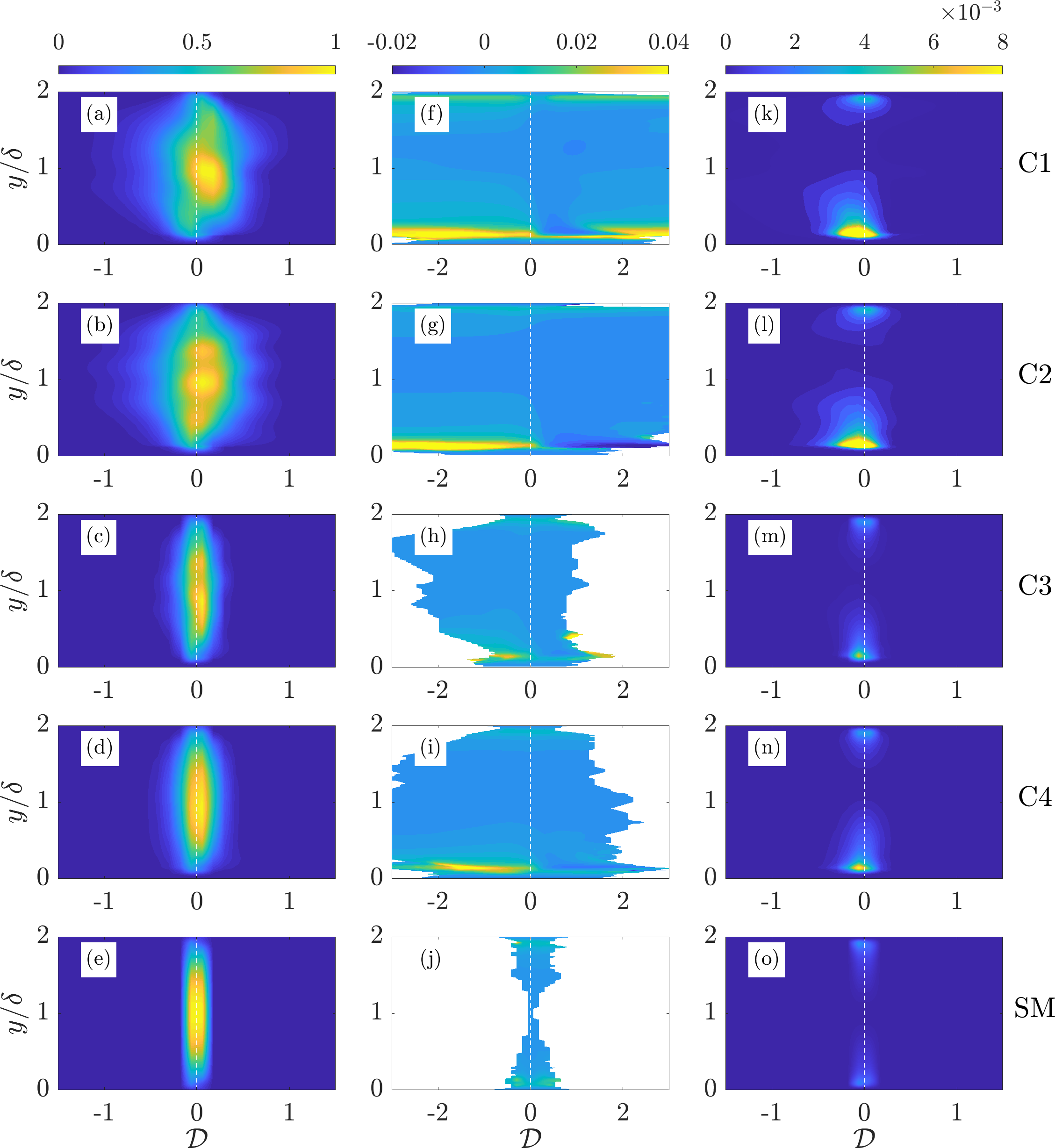}}
  \caption{Probability density functions of velocity divergence evaluated at all $y$ values (a-e), conditional expectations of TKE shear production given velocity divergence $E(\mathcal{P}_{TKE} \big| \boldsymbol{\nabla}\cdot\boldsymbol{u}=\mathcal{D})$ (f-j), and their respective products (k-o). Subplots (a-e) are normalized to yield the maximum value of 1.  }
\label{fig5:Conditional_tot}
\end{figure}

Figures \ref{fig5:Conditional_tot} (a-e) show the  probability density function (p.d.f.)  of $\boldsymbol{\nabla}\cdot\boldsymbol{u}$, evaluated at each $y$-location and normalized to yield the maximum value of 1. There $\mathcal{D}$ denotes  values taken by  $\boldsymbol{\nabla}\cdot\boldsymbol{u}$. 
{ To calculate the p.d.f. at all $y$ elevations,  the bin width is chosen as a constant 0.2 in reference units, with $(\boldsymbol{\nabla}\cdot\boldsymbol{u})_{\max}=40$, $(\boldsymbol{\nabla}\cdot\boldsymbol{u})_{\min}=-40$.
}
For the {2D roughness cases} (C1 and C2),  velocity divergence values far from the walls scatter towards much larger magnitudes than those near the wall, while they remain in the vicinity of zero for the 3D and smooth-wall cases. This is a quantitative comparison showing the more significant compressibility effects in  the 2D roughness cases  owing to the strong oblique shock waves away from the walls.  It also confirms that the 3D roughnesses  do not induce  strong compressibility effects in the flow, similar to what happens in the smooth-wall case.

Next, the TKE production conditionally averaged based on velocity divergence,  $E(\mathcal{P}_{TKE} \big| \boldsymbol{\nabla}\cdot\boldsymbol{u}=\mathcal{D})$, evaluated at each $y$, is shown in figures \ref{fig5:Conditional_tot} (f-j). In all cases, regions with large magnitudes of $\mathcal{D}$ contribute significantly to $\mathcal{P}_{TKE}$, both near and far from the wall. This is prominent in the 2D roughness cases   and, to a lower extent, in the 3D-roughness and smooth-wall cases. 
Although high-magnitude $\mathcal{D}$ events are associated with significant $\mathcal{P}_{TKE}$, their probability of occurrence is low according to the   p.d.f. of $\boldsymbol{\nabla}\cdot\boldsymbol{u}$. To assess the distribution of the actual amount of $\mathcal{P}_{TKE}$ attributed to regions of different $\boldsymbol{\nabla}\cdot\boldsymbol{u}$ values, the product between p.d.f. of $\boldsymbol{\nabla}\cdot\boldsymbol{u}$ and  $E(\mathcal{P}_{TKE} \big| \boldsymbol{\nabla}\cdot\boldsymbol{u}=\mathcal{D})$ is plotted in figures \ref{fig5:Conditional_tot}(k-o),  for each $\mathcal{D}$ and  at each $y$.
Results show that the majority of TKE production comes from low-compressibility regions, due to their high probabilities of occurrence. The 2D roughnesses lead  to larger fractions of TKE production from  negative-$\mathcal{D}$ (or compression) events.

\begin{figure}
   \centerline{\includegraphics[width=.9\textwidth,trim={0cm 0cm 0cm 0cm},clip]{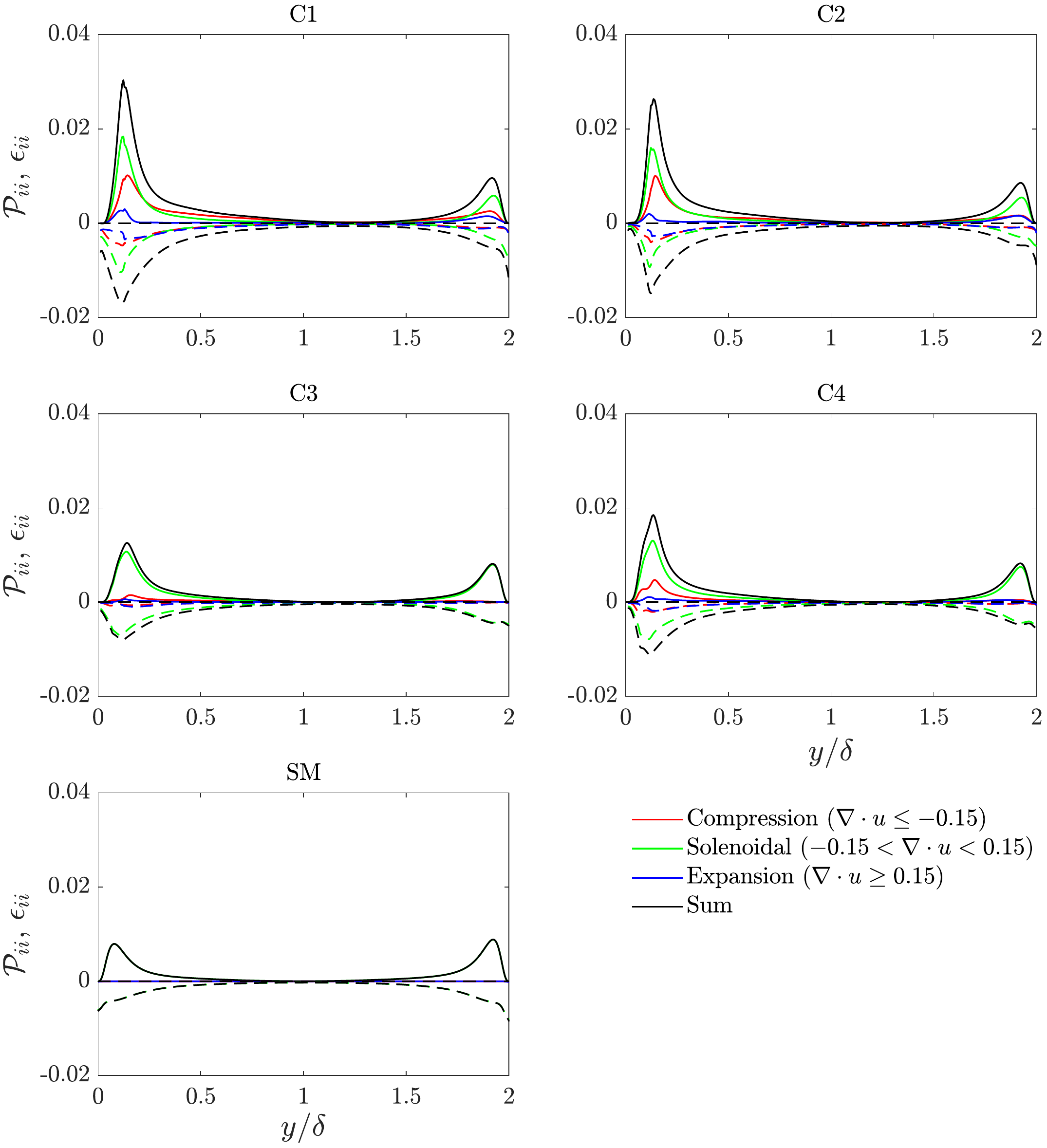}}
  \caption{Profiles of spatially averaged $\mathcal{P}_{ii}$ ($\solid$) and $\epsilon_{ii}$ ($\dashed$) conditioned on various types of compressibility. Quantities are normalized using $\rho_r$, $U_r$ and $\delta$. 
  }
\label{fig5:Conditional_P_Ep}
\end{figure}

To quantitatively compare individual contributions from regions with different types of compressibility to the overall TKE production, figure \ref{fig5:Conditional_P_Ep} shows profiles of  $\mathcal{P}_{TKE}$ and $\epsilon_{TKE}$ conditioned on three types of events: compression events (where $\frac{\boldsymbol{\nabla}\cdot\boldsymbol{u}}{U_r/\delta}\leq-0.15$), solenoidal events ($-0.15<\frac{\boldsymbol{\nabla}\cdot\boldsymbol{u}}{U_r/\delta}<0.15$) and expansion events ($\frac{\boldsymbol{\nabla}\cdot\boldsymbol{u}}{U_r/\delta}\geq 0.15$). 
{
The threshold value 0.15 is chosen from figure \ref{fig5:Conditional_tot}(e) as approximately the maximum magnitude reached in the smooth case. In other words, for the purpose of the conditional analysis, the  smooth-wall flow at the present Mach number is considered only weakly compressible (relative to the rough-wall cases).
}
For the 3D-roughnesses and smooth-wall cases, the solenoidal events are shown  responsible for almost all of the TKE production (and dissipation rate). However, for 2D surfaces  the compression events  contribute as much as 30\% on these processes near the wall, while the contribution from expansion events is about 5\%; far from the wall ($y/\delta > 0.2$), the total contribution from   compression events overtakes that from   solenoidal ones.
This indicates that the shocks,   dependent on the roughness topography, dynamically influences the turbulent flow across the channel.
{ For further understanding of how roughness geometry affects the Reynolds stress balance, future studies are needed to characterize the dilatation terms \citep[i.e. pressure dilatation and dilatation dissipation,][]{SarkarSEH91}, which  are normally  small at present Mach and Reynolds numbers in channel flow with smooth walls but may be significant on rough walls due to the present of  shocks. }

\section{Concluding remarks}\label{sec5:conclusion}

Effects of surface roughness and its topography on compressible turbulent flows were characterized based on simulations of four supersonic channel flows at $M=1.5$ and the bulk Reynolds number $\text{Re}=3000$  with top smooth walls and four different roughness geometries on the bottom walls. A baseline smooth-wall channel was also simulated. { A modified  level-set/volume-of-fluid immersed boundary method was used to impose the boundary conditions of velocity and temperature on the surface of roughness. The method was validated in terms of  mean  and turbulence statistics against a companion conformal mesh simulation.}
The four roughnesses include two 2D and two 3D sinusoidal surfaces. The surfaces shared the same peak-to-trough height (of 10\% channel half height) but differed in the surface wavelength. 

Results showed significant modifications of turbulence across the channel by the roughness. Roughness generates a distribution of oblique shocks in connection to the roughness geometry. These shocks are much stronger on the 2D roughnesses; they reach across the channel,  reflected back from the smooth-wall side, and interact to form shock diamonds.  Such strong shocks and pattern are absent in the smooth-wall channel.

The strong shocks generated by the 2D roughnesses  lead to stronger irreversible heat generation and higher temperature values in the bulk of the channel, than in the 3D-roughness cases. 
Roughness on one side of the channel enhances TKE production on both sides. Conditional analyses based on local compressibility showed that, although high-compressibility regions are associated with  significantly enhanced local TKE production, the probability of occurrence of such regions is low and dependent on the roughness geometry. For the 2D roughness cases up to 30\% of  TKE production and dissipation near the rough wall (as well as most of the outer-layer production  and dissipation) occur in compression regions. However, for the 3D-roughness and smooth-wall cases these processes are almost all associated with relatively solenoidal events.

This work identifies the mechanism through which wall roughness and its texture  affect a compressible turbulence. It demonstrates that in supersonic channels the roughness effects can propagate throughout the boundary layer. In this case, the  outer layer similarity established for incompressible flows does not apply.   { This work focuses on flow statistics. To fully characterize roughness effects in compressible flows, future studies on spectral and structural effects,  scaling and variable density effects are needed. }

\section*{Acknowledgements}
Computational support was provided by Michigan State University’s Institute for Cyber-Enabled Research. 

\section*{Declaration of Interests}
The authors report no conflict of interest.

\bibliographystyle{jfm}

\bibliography{biblio}

\begin{thebibliography}{73}
\expandafter\ifx\csname natexlab\endcsname\relax\def\natexlab#1{#1}\fi
\def\au#1{#1} \def\ed#1{#1} \def\yr#1{#1}\def\at#1{#1}\def\jt#1{\textit{#1}}
  \def\bt#1{#1}\def\bvol#1{\textbf{#1}} \def\vol#1{#1} \def\pg#1{#1}
  \def\publ#1{#1}\def\arxiv#1{#1}\def\org#1{#1}\def\st#1{\textit{#1}}

\bibitem[Aghaei~Jouybari {\em et~al.\/}(2021)Aghaei~Jouybari, Yuan, Brereton \&
  Murillo]{Aghaei-Jouybari2021}
{\sc \au{Aghaei~Jouybari, Mostafa}, \au{Yuan, Junlin}, \au{Brereton, Giles~J.}
  \& \au{Murillo, Michael~S.}} \yr{2021}  \at{Data-driven prediction of the
  equivalent sand-grain height in rough-wall turbulent flows}.  \jt{Journal of
  Fluid Mechanics}  \bvol{912},  \pg{A8}.

\bibitem[Anderson(1990)]{Anderson90}
{\sc \au{Anderson, J.D.}} \yr{1990} {\em Modern compressible flow: with
  historical perspective\/}, ,  \vol{vol.~12}.  \publ{McGraw-Hill New York}.

\bibitem[Bernardini {\em et~al.\/}(2012)Bernardini, Pirozzoli \&
  P.]{Bernardini12}
{\sc \au{Bernardini, M.}, \au{Pirozzoli, S.} \& \au{P., Orlandi.}} \yr{2012}
  \at{Compressibility effects on roughness-induced boundary layer transition}.
  \jt{Int. J. Heat Fluid Flow}  \bvol{35},  \pg{45 -- 51}.

\bibitem[Braslow \& Knox(1958)]{Braslow58}
{\sc \au{Braslow, A.~L.} \& \au{Knox, E.~C.}} \yr{1958}  \bt{Simplified method
  for determination of critical height of distributed roughness particles for
  boundary-layer transition at {M}ach numbers from 0 to 5}. {\em Tech. Rep.\/}.
   \org{National Advisory Committee for Aeronautics}.

\bibitem[Busse {\em et~al.\/}(2017)Busse, Thakkar \& Sandham]{BusseBTS17}
{\sc \au{Busse, A.}, \au{Thakkar, M.} \& \au{Sandham, N.~D.}} \yr{2017}
  \at{{R}eynolds-number dependence of the near-wall flow over irregular rough
  surfaces}.  \jt{J. Fluid Mech.}  \bvol{810},  \pg{196--224}.

\bibitem[Chaudhuri {\em et~al.\/}(2011)Chaudhuri, Hadjadj \&
  Chinnayya]{Chaudhuri11}
{\sc \au{Chaudhuri, A.}, \au{Hadjadj, A.} \& \au{Chinnayya, A.}} \yr{2011}
  \at{On the use of immersed boundary methods for shock/obstacle interactions}.
   \jt{J. Comput. Phys.}  \bvol{230}~(5),  \pg{1731 -- 1748}.

\bibitem[Chong {\em et~al.\/}(1990)Chong, Perry \& Cantwell]{Chong90}
{\sc \au{Chong, M.~S.}, \au{Perry, A.~E.} \& \au{Cantwell, B.~J.}} \yr{1990}
  \at{A general classification of three‐dimensional flow fields}.
  \jt{Physics of Fluids A: Fluid Dynamics}  \bvol{2}~(5),  \pg{765--777}.

\bibitem[Coleman {\em et~al.\/}(1995)Coleman, Kim \& Moser]{ColemanKM95}
{\sc \au{Coleman, G.~N.}, \au{Kim, J.} \& \au{Moser, R.~D.}} \yr{1995}  \at{{A
  numerical study of turbulent supersonic isothermal-wall channel flow}}.
  \jt{J. Fluid Mech.}  \bvol{305},  \pg{159--183}.

\bibitem[Ekoto {\em et~al.\/}(2008)Ekoto, W.~Bowersox, Beutner \&
  Goss]{Ekoto08}
{\sc \au{Ekoto, I.~W.}, \au{W.~Bowersox, R.~D.}, \au{Beutner, T.} \& \au{Goss,
  L.~P.}} \yr{2008}  \at{Supersonic boundary layers with periodic surface
  roughness}.  \jt{AIAA Journal}  \bvol{46},  \pg{486 -- 497}.

\bibitem[Fadlun {\em et~al.\/}(2000)Fadlun, Verzicco, Orlandi \&
  Mohd-Yusof]{FadlunVOM00}
{\sc \au{Fadlun, E.~A.}, \au{Verzicco, R.}, \au{Orlandi, P.} \& \au{Mohd-Yusof,
  J.}} \yr{2000}  \at{Combined immersed-boundary finite-difference methods for
  three-dimensional complex flow simulations}.  \jt{J. Comput. Phys.}
  \bvol{161},  \pg{35--60}.

\bibitem[Flack(2018)]{FlackF18a}
{\sc \au{Flack, K.~A.}} \yr{2018}  \at{Moving beyond {M}oody}.  \jt{J. Fluid
  Mech.}  \bvol{842},  \pg{1--4}.

\bibitem[Flack {\em et~al.\/}(2007)Flack, Schultz \& Connelly]{FlackSC07}
{\sc \au{Flack, K.~A.}, \au{Schultz, M.~P.} \& \au{Connelly, J.~S.}} \yr{2007}
  \at{{Examination of a critical roughness height for outer layer similarity}}.
   \jt{Phys.\ Fluids}  \bvol{19},  \pg{095104}.

\bibitem[Forooghi {\em et~al.\/}(2017)Forooghi, Stroh, Magagnato, Jakirlic \&
  Frohnapfel]{ForooghiFSMJF17a}
{\sc \au{Forooghi, P.}, \au{Stroh, A.}, \au{Magagnato, F.}, \au{Jakirlic, S.}
  \& \au{Frohnapfel, B.}} \yr{2017}  \at{Toward a universal roughness
  correlation}.  \jt{J. Fluids Eng.}  \bvol{139},  \pg{121201--1--12}.

\bibitem[Foysi {\em et~al.\/}(2004)Foysi, Sarkar \& Friedrich]{Foysi04}
{\sc \au{Foysi, H.}, \au{Sarkar, S.} \& \au{Friedrich, R.}} \yr{2004}
  \at{Compressibility effects and turbulence scalings in supersonic channel
  flow}.  \jt{Journal of Fluid Mechanics}  \bvol{509},  \pg{207--216}.

\bibitem[Ghias {\em et~al.\/}(2007)Ghias, Mittal \& Dong]{Ghias07}
{\sc \au{Ghias, R.}, \au{Mittal, R.} \& \au{Dong, H.}} \yr{2007}  \at{A sharp
  interface immersed boundary method for compressible viscous flows}.  \jt{J.
  Comput. Phys.}  \bvol{225},  \pg{528 -- 553}.

\bibitem[Gibou {\em et~al.\/}(2018)Gibou, Fedkiw \& Osher]{Gibou18}
{\sc \au{Gibou, F.}, \au{Fedkiw, R.} \& \au{Osher, S.}} \yr{2018}  \at{A review
  of level-set methods and some recent applications}.  \jt{J. Comput. Phys.}
  \bvol{353},  \pg{82 -- 109}.

\bibitem[Guarini {\em et~al.\/}(2000)Guarini, Moser, Shariff \&
  Wray]{Guarini2000}
{\sc \au{Guarini, S.E.}, \au{Moser, R.D.}, \au{Shariff, K.} \& \au{Wray, A.}}
  \yr{2000}  \at{Direct numerical simulation of a supersonic turbulent boundary
  layer at mach 2.5}.  \jt{Journal of Fluid Mechanics}  \bvol{414},
  \pg{1--33}.

\bibitem[Jammalamadaka \& Jaberi(2015)]{Jammalamadaka15}
{\sc \au{Jammalamadaka, A.} \& \au{Jaberi, F.A.}} \yr{2015}  \at{Subgrid-scale
  turbulence in shock--boundary layer flows}.  \jt{Theoretical and
  Computational Fluid Dynamics}  \bvol{29}~(1),  \pg{29--54}.

\bibitem[Jammalamadaka {\em et~al.\/}(2013)Jammalamadaka, Li \&
  Jaberi]{Jammalamadaka13}
{\sc \au{Jammalamadaka, A.}, \au{Li, Z.} \& \au{Jaberi, F.A.}} \yr{2013}
  \at{Subgrid-scale models for large-eddy simulations of shock-boundary-layer
  interactions}.  \jt{AIAA journal}  \bvol{51}~(5),  \pg{1174--1188}.

\bibitem[Jammalamadaka {\em et~al.\/}(2014)Jammalamadaka, Li \&
  Jaberi]{Jammalamadaka14}
{\sc \au{Jammalamadaka, A.}, \au{Li, Z.} \& \au{Jaberi, F.A.}} \yr{2014}
  \at{Numerical investigations of shock wave interactions with a supersonic
  turbulent boundary layer}.  \jt{Physics of Fluids}  \bvol{26}~(5),
  \pg{056101}.

\bibitem[Ji {\em et~al.\/}(2006)Ji, Yuan \& Chung]{Ji06}
{\sc \au{Ji, Y.}, \au{Yuan, K.} \& \au{Chung, J.~N.}} \yr{2006}  \at{Numerical
  simulation of wall roughness on gaseous flow and heat transfer in a
  microchannel}.  \jt{Int. J. Heat Mass Transf.}  \bvol{49}~(7),  \pg{1329 --
  1339}.

\bibitem[Jim{\'e}nez(2004)]{Jimenez04}
{\sc \au{Jim{\'e}nez, J.}} \yr{2004}  \at{Turbulent flows over rough walls}.
  \jt{Annu. Rev. Fluid Mech.}  \bvol{36},  \pg{173--196}.

\bibitem[Lagha {\em et~al.\/}(2011)Lagha, Kim, Eldredge \& Zhong]{Lagha11}
{\sc \au{Lagha, M.}, \au{Kim, J.}, \au{Eldredge, J.D.} \& \au{Zhong, X.}}
  \yr{2011}  \at{A numerical study of compressible turbulent boundary layers}.
  \jt{Physics of Fluids}  \bvol{23}~(1),  \pg{015106}.

\bibitem[Latin(1998)]{Latin98}
{\sc \au{Latin, R.~M.}} \yr{1998}  \at{The influence of surface roughness on
  supersonic high {R}eynolds number turbulent boundary layer flow}. PhD thesis,
  School of Engineering of the Air Force Institute of Technology Air
  University.

\bibitem[Latin \& W.~Bowersox(2000)]{Latin00}
{\sc \au{Latin, R.~M.} \& \au{W.~Bowersox, R.~D.}} \yr{2000}  \at{Flow
  properties of a supersonic turbulent boundary layer with wall roughness}.
  \jt{AIAA Journal}  \bvol{38},  \pg{1804 -- 1821}.

\bibitem[Latin \& W.~Bowersox(2002)]{Latin02}
{\sc \au{Latin, R.~M.} \& \au{W.~Bowersox, R.~D.}} \yr{2002}  \at{Temporal
  turbulent flow structure for supersonic rough-wall boundary layers}.
  \jt{AIAA Journal}  \bvol{40},  \pg{832 -- 841}.

\bibitem[Leonardi {\em et~al.\/}(2007)Leonardi, Orlandi \& Antonia]{Leonardi07}
{\sc \au{Leonardi, S.}, \au{Orlandi, P.} \& \au{Antonia, R.~A.}} \yr{2007}
  \at{Properties of d- and k-type roughness in a turbulent channel flow}.
  \jt{Phys. Fluids}  \bvol{19}~(12),  \pg{125101}.

\bibitem[Li {\em et~al.\/}(2008)Li, Jaberi \& Shih]{Li08}
{\sc \au{Li, Z.}, \au{Jaberi, F.A.} \& \au{Shih, T.}} \yr{2008}  \at{A hybrid
  lagrangian--eulerian particle-level set method for numerical simulations of
  two-fluid turbulent flows}.  \jt{International Journal for Numerical Methods
  in Fluids}  \bvol{56}~(12),  \pg{2271--2300}.

\bibitem[Li \& Jaberi(2012)]{LiJ12}
{\sc \au{Li, Z.} \& \au{Jaberi, F.~A.}} \yr{2012}  \at{A high-order finite
  difference method for numerical simulations of supersonic turbulent flows}.
  \jt{Int. J. Numer. Meth. Fl.}  \bvol{68}~(6),  \pg{740--766}.

\bibitem[Luo {\em et~al.\/}(2017)Luo, Mao, Zhuang, Fan \& Haugen]{Luo17}
{\sc \au{Luo, K.}, \au{Mao, C.}, \au{Zhuang, Z.}, \au{Fan, J.} \& \au{Haugen,
  N.E.L}} \yr{2017}  \at{{A ghost-cell immersed boundary method for the
  simulations of heat transfer in compressible flows under different boundary
  conditions Part-II: Complex geometries}}.  \jt{Int. J. Heat Mass Tran.}
  \bvol{104},  \pg{98 -- 111}.

\bibitem[Ma {\em et~al.\/}(2021)Ma, Alame \& Mahesh]{Ma2021}
{\sc \au{Ma, R.}, \au{Alame, K.} \& \au{Mahesh, K.}} \yr{2021}  \at{Direct
  numerical simulation of turbulent channel flow over random rough surfaces}.
  \jt{J. Fluid Mech.}  \bvol{908},  \pg{A40}.

\bibitem[Mangavelli {\em et~al.\/}(2021)Mangavelli, Yuan \&
  Brereton]{Mangavelli21}
{\sc \au{Mangavelli, S.C.}, \au{Yuan, J.} \& \au{Brereton, G.J.}} \yr{2021}
  \at{Effects of surface roughness topography in transient channel flows}.
  \jt{Journal of Turbulence}  \pg{pp. 1--27}.

\bibitem[Marusic {\em et~al.\/}(2010)Marusic, McKeon, Monkewitz, Nagib, Smits
  \& Sreenivasan]{Marusic10}
{\sc \au{Marusic, I.}, \au{McKeon, B.~J.}, \au{Monkewitz, P.~A.}, \au{Nagib,
  H.~M.}, \au{Smits, A.~J.} \& \au{Sreenivasan, K.~R.}} \yr{2010}
  \at{Wall-bounded turbulent flows at high {R}eynolds numbers: Recent advances
  and key issues}.  \jt{Phys. Fluids}  \bvol{22}~(6),  \pg{065103}.

\bibitem[Mejia-Alvarez \& Christensen(2010)]{Mejia-AlvarezC10}
{\sc \au{Mejia-Alvarez, R.} \& \au{Christensen, K.~T.}} \yr{2010}
  \at{Low-order representations of irregular surface roughness and their impact
  on a turbulent boundary layer}.  \jt{Phys.\ Fluids}  \bvol{22},
  \pg{015106--1--20}.

\bibitem[Mittal \& Iaccarino(2005)]{MittalI05}
{\sc \au{Mittal, R.} \& \au{Iaccarino, G.}} \yr{2005}  \at{Immersed boundary
  methods}.  \jt{Annu.\ Rev.\ Fluid Mech.}  \bvol{37},  \pg{239--261}.

\bibitem[Modesti {\em et~al.\/}(2022)Modesti, Sathyanarayana, Salvadore \&
  Bernardini]{Modesti22}
{\sc \au{Modesti, D.}, \au{Sathyanarayana, S.}, \au{Salvadore, F.} \&
  \au{Bernardini, M.}} \yr{2022}  \at{Direct numerical simulation of supersonic
  turbulent flows over rough surfaces}.  \jt{Journal of Fluid Mechanics}
  \bvol{942},  \pg{A44}.

\bibitem[Morkovin(1962)]{Morkovin62}
{\sc \au{Morkovin, M.V.}} \yr{1962} Effects of compressibility on turbulent
  flows.  \bt{In {\em M{\'e}canique de la Turbulence\/}},  \pg{pp. 367--380}.

\bibitem[Muppidi \& Mahesh(2012)]{Muppidi12}
{\sc \au{Muppidi, S.} \& \au{Mahesh, K.}} \yr{2012}  \at{Direct numerical
  simulations of roughness-induced transition in supersonic boundary layers}.
  \jt{J. Fluid Mech.}  \bvol{693},  \pg{28--56}.

\bibitem[Napoli {\em et~al.\/}(2008)Napoli, Armenio \&
  De~Marchis]{NapoliNAD08a}
{\sc \au{Napoli, E.}, \au{Armenio, V.} \& \au{De~Marchis, M.}} \yr{2008}
  \at{The effect of the slope of irregularly distributed roughness elements on
  turbulent wall-bounded flows}.  \jt{J. Fluid Mech.}  \bvol{613},
  \pg{385--394}.

\bibitem[Nikuradse(1933)]{Nikuradse33e}
{\sc \au{Nikuradse, J.}} \yr{1933}  \at{Laws of flow in rough pipes}.  \jt{NACA
  Technical Memorandum 1292} .

\bibitem[Radeztsky {\em et~al.\/}(1999)Radeztsky, Reibert \&
  Saric]{Radeztsky99}
{\sc \au{Radeztsky, R.~H.}, \au{Reibert, M.~S.} \& \au{Saric, W.~S.}} \yr{1999}
   \at{Effect of isolated micron-sized roughness on transition in swept-wing
  flows}.  \jt{AIAA Journal}  \bvol{37},  \pg{1370 -- 1377}.

\bibitem[Raupach {\em et~al.\/}(1991)Raupach, Antonia \&
  Rajagopalan]{RaupachAR91}
{\sc \au{Raupach, M.~R.}, \au{Antonia, R.~A.} \& \au{Rajagopalan, S.}}
  \yr{1991}  \at{Rough-wall boundary layers}.  \jt{Appl. Mech. Rev.}
  \bvol{44},  \pg{1--25}.

\bibitem[Raupach \& Shaw(1982)]{RaupachS82}
{\sc \au{Raupach, M.~R.} \& \au{Shaw, R.~H.}} \yr{1982}  \at{Averaging
  procedures for flow within vegetation canopies}.  \jt{Bound.-Lay.\ Meteorol.}
   \bvol{22},  \pg{79--90}.

\bibitem[Reda(2002)]{Reda02}
{\sc \au{Reda, D.~C.}} \yr{2002}  \at{Review and synthesis of
  roughness-dominated transition correlations for reentry applications}.
  \jt{J. Spacecr. Rockets}  \bvol{39},  \pg{161--167}.

\bibitem[Reda {\em et~al.\/}(2008)Reda, Wilder, Bogdanoff \& Prabhu]{Reda08}
{\sc \au{Reda, D.~C.}, \au{Wilder, M.~C.}, \au{Bogdanoff, D.~W.} \& \au{Prabhu,
  D.~K.}} \yr{2008}  \at{Transition experiments on blunt bodies with
  distributed roughness in hypersonic free flight}.  \jt{J. Spacecr. Rockets}
  \bvol{45},  \pg{210 -- 215}.

\bibitem[Reshotko \& Tumin(2004)]{Reshotko04}
{\sc \au{Reshotko, E.} \& \au{Tumin, A.}} \yr{2004}  \at{Role of transient
  growth in roughness-induced transition}.  \jt{AIAA Journal}  \bvol{42},
  \pg{766 -- 770}.

\bibitem[Sarkar {\em et~al.\/}(1991)Sarkar, Erlebacher, Hussaini \&
  Kreiss]{SarkarSEH91}
{\sc \au{Sarkar, S.}, \au{Erlebacher, G.}, \au{Hussaini, M.~Y.} \& \au{Kreiss,
  H.~O.}} \yr{1991}  \at{The analysis and modelling of dilatational terms in
  compressible turbulence}.  \jt{J. Fluid Mech.}  \bvol{227},  \pg{473--493}.

\bibitem[Schneider(2008)]{Schneider08}
{\sc \au{Schneider, S.~P.}} \yr{2008}  \at{Effects of roughness on hypersonic
  boundary-layer transition}.  \jt{J. Spacecr. Rockets}  \bvol{45},
  \pg{193--209}.

\bibitem[Schultz \& Flack(2007)]{SchultzF07}
{\sc \au{Schultz, M.~P.} \& \au{Flack, K.~A.}} \yr{2007}  \at{{The rough-wall
  turbulent boundary layer from the hydraulically smooth to the fully rough
  regime}}.  \jt{J. Fluid Mech.}  \bvol{580},  \pg{381--405}.

\bibitem[Scotti(2006)]{Scotti06}
{\sc \au{Scotti, A.}} \yr{2006}  \at{Direct numerical simulation of turbulent
  channel flows with boundary roughened with virtual sandpaper}.  \jt{Phys.\
  Fluids}  \bvol{18},  \pg{031701--1---4}.

\bibitem[Shen {\em et~al.\/}(2020)Shen, Yuan \& Phanikumar]{Shen20}
{\sc \au{Shen, G.}, \au{Yuan, J.} \& \au{Phanikumar, M.S.}} \yr{2020}
  \at{Direct numerical simulations of turbulence and hyporheic mixing near
  sediment--water interfaces}.  \jt{Journal of Fluid Mechanics}  \bvol{892}.

\bibitem[Sussman {\em et~al.\/}(1999)Sussman, Almgren, Bell, Colella, Howell \&
  Welcome]{Sussman99}
{\sc \au{Sussman, M.}, \au{Almgren, A.S.}, \au{Bell, J.B.}, \au{Colella, P.},
  \au{Howell, L.H.} \& \au{Welcome, M.L.}} \yr{1999}  \at{An adaptive level set
  approach for incompressible two-phase flows}.  \jt{J. Comput. Phys.}
  \bvol{148}~(1),  \pg{81--124}.

\bibitem[Sussman {\em et~al.\/}(1994)Sussman, Smereka \& Osher]{Sussman94}
{\sc \au{Sussman, M.}, \au{Smereka, P.} \& \au{Osher, S.}} \yr{1994}  \at{A
  level set approach for computing solutions to incompressible two-phase flow}.
   \jt{J. Comput. Phys.}  \bvol{114}~(1),  \pg{146 -- 159}.

\bibitem[Talapatra \& Katz(2012)]{TalapatraK12}
{\sc \au{Talapatra, S.} \& \au{Katz, J.}} \yr{2012}  \at{{Coherent structures
  in the inner part of a rough-wall channel flow resolved using holographic
  PIV}}.  \jt{J. Fluid Mech.}  \bvol{711},  \pg{161--170}.

\bibitem[Thakkar {\em et~al.\/}(2018)Thakkar, Busse \& Sandham]{ThakkarTBS18a}
{\sc \au{Thakkar, M.}, \au{Busse, A.} \& \au{Sandham, N.~D.}} \yr{2018}
  \at{Direct numerical simulation of turbulent channel flow over a surrogate
  for {Nikuradse-type} roughness}.  \jt{Journal of Fluid Mechanics}
  \bvol{837},  \pg{R1}.

\bibitem[Tian {\em et~al.\/}(2017)Tian, Jaberi, Li \& Livescu]{Tian17}
{\sc \au{Tian, Y.}, \au{Jaberi, F.A.}, \au{Li, Z.} \& \au{Livescu, D.}}
  \yr{2017}  \at{Numerical study of variable density turbulence interaction
  with a normal shock wave}.  \jt{Journal of Fluid Mechanics}  \bvol{829},
  \pg{551--588}.

\bibitem[Tian {\em et~al.\/}(2019)Tian, Jaberi \& Livescu]{Tian19}
{\sc \au{Tian, Y.}, \au{Jaberi, F.A.} \& \au{Livescu, D.}} \yr{2019}
  \at{Density effects on post-shock turbulence structure and dynamics}.
  \jt{Journal of Fluid Mechanics}  \bvol{880},  \pg{935--968}.

\bibitem[Townsend(1976)]{Townsend76}
{\sc \au{Townsend, A.~A.}} \yr{1976} {\em The structure of turbulent shear
  flow\/}.  \publ{Cambridge University Press}.

\bibitem[Trettel \& Larsson(2016)]{Trettel16}
{\sc \au{Trettel, A.} \& \au{Larsson, J.}} \yr{2016}  \at{Mean velocity scaling
  for compressible wall turbulence with heat transfer}.  \jt{Physics of Fluids}
   \bvol{28}~(2),  \pg{026102},  \arxiv{arXiv:
  https://doi.org/10.1063/1.4942022}.

\bibitem[Tyson \& Sandham(2013)]{Tyson13}
{\sc \au{Tyson, C.J.} \& \au{Sandham, N.D.}} \yr{2013}  \at{Numerical
  simulation of fully-developed compressible flows over wavy surfaces}.
  \jt{Int. J. Heat Fluid Flow}  \bvol{41},  \pg{2 -- 15}.

\bibitem[Van~Driest(1951)]{Van-Driest51}
{\sc \au{Van~Driest, E.R.}} \yr{1951}  \at{Turbulent boundary layer in
  compressible fluids}.  \jt{Journal of the Aeronautical Sciences}
  \bvol{18}~(3),  \pg{145--160},  \arxiv{arXiv:
  https://doi.org/10.2514/8.1895}.

\bibitem[Vitturi {\em et~al.\/}(2007)Vitturi, Ongaro, Neri, Salvetti \&
  Beux]{Vitturi07}
{\sc \au{Vitturi, M. D.~M.}, \au{Ongaro, T.~E.}, \au{Neri, A.}, \au{Salvetti,
  M.~V.} \& \au{Beux, F.}} \yr{2007}  \at{An immersed boundary method for
  compressible multiphase flows: application to the dynamics of pyroclastic
  density currents}.  \jt{Comput. Geosci.}  \bvol{11},  \pg{183--198}.

\bibitem[Volino {\em et~al.\/}(2011)Volino, Schultz \& Flack]{VolinoVSF11}
{\sc \au{Volino, R.~J.}, \au{Schultz, M.~P.} \& \au{Flack, K.~A.}} \yr{2011}
  \at{Turbulence structure in boundary layers over periodic two- and
  three-dimensional roughness}.  \jt{J. Fluid Mech.}  \bvol{676},
  \pg{172--190}.

\bibitem[Volpiani {\em et~al.\/}(2020)Volpiani, Iyer, Pirozzoli \&
  Larsson]{Volpiani20}
{\sc \au{Volpiani, P.S.}, \au{Iyer, P.S.}, \au{Pirozzoli, S.} \& \au{Larsson,
  J.}} \yr{2020}  \at{Data-driven compressibility transformation for turbulent
  wall layers}.  \jt{Phys. Rev. Fluids}  \bvol{5},  \pg{052602}.

\bibitem[Vyas {\em et~al.\/}(2019)Vyas, Yoder \& Gaitonde]{Vyas19}
{\sc \au{Vyas, M.A.}, \au{Yoder, D.A.} \& \au{Gaitonde, D.V.}} \yr{2019}
  \at{Reynolds-stress budgets in an impinging shock-wave/boundary-layer
  interaction}.  \jt{AIAA Journal}  \bvol{57}~(11),  \pg{4698--4714}.

\bibitem[Wang {\em et~al.\/}(2017)Wang, Currao, Han, J., Young \& Tian]{Wang17}
{\sc \au{Wang, L.}, \au{Currao, G.M.D.}, \au{Han, F.}, \au{J., Neely.~A.},
  \au{Young, J.} \& \au{Tian, F.B.}} \yr{2017}  \at{An immersed boundary method
  for fluid-structure interaction with compressible multiphase flows}.  \jt{J.
  Comput. Phys.}  \bvol{346},  \pg{131 -- 151}.

\bibitem[Womack {\em et~al.\/}(2022)Womack, Volino, Meneveau \&
  Schultz]{Womack22}
{\sc \au{Womack, K.M.}, \au{Volino, R.J.}, \au{Meneveau, C.} \& \au{Schultz,
  M.P.}} \yr{2022}  \at{Turbulent boundary layer flow over regularly and
  irregularly arranged truncated cone surfaces}.  \jt{Journal of Fluid
  Mechanics}  \bvol{933},  \pg{A38}.

\bibitem[Yang {\em et~al.\/}(2015)Yang, Sadique, Mittal \& Meneveau]{Yang15}
{\sc \au{Yang, X.I.A.}, \au{Sadique, J.}, \au{Mittal, R.} \& \au{Meneveau, C.}}
  \yr{2015}  \at{Integral wall model for large eddy simulations of wall-bounded
  turbulent flows}.  \jt{Physics of Fluids}  \bvol{27}~(2),  \pg{025112}.

\bibitem[Yuan \& Aghaei~Jouybari(2018)]{YuanA18}
{\sc \au{Yuan, J.} \& \au{Aghaei~Jouybari, M.}} \yr{2018}  \at{Topographical
  effects of roughness on turbulence statistics in roughness sublayer}.
  \jt{Phys. Rev. Fluids}  \bvol{3},  \pg{114603--19}.

\bibitem[Yuan {\em et~al.\/}(2019)Yuan, Mishra, Brereton, Iaccarino \&
  Vartdal]{YuanYMBIV19a}
{\sc \au{Yuan, J.}, \au{Mishra, A.~A.}, \au{Brereton, G.~J.}, \au{Iaccarino,
  G.} \& \au{Vartdal, M.}} \yr{2019}  \at{Single-point structure tensors in
  turbulent channel flows with smooth and wavy walls}.  \jt{Phys. Fluids}
  \bvol{31},  \pg{125115--1--15}.

\bibitem[Yuan \& Piomelli(2014)]{YuanP14b}
{\sc \au{Yuan, J.} \& \au{Piomelli, U.}} \yr{2014}  \at{Roughness effects on
  the {Reynolds} stress budgets in near-wall turbulence}.  \jt{J. Fluid Mech.}
  \bvol{760},  \pg{R1}.

\bibitem[Yuan \& Piomelli(2015)]{YuanYP15}
{\sc \au{Yuan, J.} \& \au{Piomelli, U.}} \yr{2015}  \at{Numerical simulation of
  a spatially developing accelerating boundary layer over roughness}.  \jt{J.
  Fluid Mech.}  \bvol{780},  \pg{192--214}.

\bibitem[Yuan \& Zhong(2018)]{YuanZ18}
{\sc \au{Yuan, R.} \& \au{Zhong, C.}} \yr{2018}  \at{An immersed-boundary
  method for compressible viscous flows and its application in the gas-kinetic
  {BGK} scheme}.  \jt{Appl. Math. Model.}  \bvol{55},  \pg{417 -- 446}.

\end{thebibliography}

\end{document}